\documentclass[twocolumn,aps,prl,amsmath,superscriptaddress,longbibliography,notitlepage]{revtex4}
\usepackage{times}
\usepackage{txfonts}
\usepackage{amssymb}
\usepackage{mathrsfs}
\usepackage{graphicx}
\usepackage{float}
\usepackage[caption=false]{subfig}
\usepackage[colorlinks,bookmarks=true,citecolor=blue,linkcolor=blue,urlcolor=blue]{hyperref}
\usepackage{verbatim}
\usepackage{epstopdf}
\usepackage[normalem]{ulem}
\usepackage{verbatim}
\usepackage{xcolor}
\usepackage{bm}
\usepackage{physics}
\usepackage{ulem}
\newcommand{\clr}{\color{red!75!black}}

\renewcommand{\Re}{\operatorname{Re}}
\renewcommand{\Im}{\operatorname{Im}}

\begin{document}
\title{Algebraic States in Continuum in d\textgreater 1 Dimensional Non-Hermitian Systems}
\author{Ao Yang}
\affiliation{Beijing National Laboratory for Condensed Matter Physics, and Institute of Physics, Chinese Academy of Sciences, Beijing 100190, China}
\affiliation{University of Chinese Academy of Sciences, Beijing 100049, China}

\author{Kai Zhang}
\affiliation{Department of Physics, University of Michigan Ann Arbor, Ann Arbor, Michigan, 48109, United States}

\author{Chen Fang}
\email{cfang@iphy.ac.cn}
\affiliation{Beijing National Laboratory for Condensed Matter Physics, and Institute of Physics, Chinese Academy of Sciences, Beijing 100190, China}
\affiliation{Kavli Institute for Theoretical Sciences, Chinese Academy of Sciences, Beijing 100190, China}

\begin{abstract}
	We report the existence of algebraically localized eigenstates embedded within the continuum spectrum of 2D non-Hermitian systems with a single impurity. 
	These modes, which we term algebraic states in continuum (AICs), decay algebraically as $1/|r|$ from the impurity site, and their energies lie within the bulk continuum spectrum under periodic boundary conditions. 
	We analytically derive the threshold condition for the impurity strength required to generate such states. 
	Remarkably, AICs are forbidden in Hermitian systems and in 1D non-Hermitian systems, making them unique to non-Hermitian systems in two and higher dimensions. 
	To detect AICs, we introduce a local density of states as an experimental observable, which is readily accessible in photonic/acoustic platforms. 
\end{abstract}

\maketitle

\emph{{\clr Introduction}.---}~The distinction between localized and extended states, which determines essential properties such as conductivity, is fundamental to understanding diverse phases of matter~\cite{Philip2012,Anderson1958,Evers2008}. 
In Hermitian systems, localized states typically have discrete energies and are separated from the continuum of extended scattering states~\cite{Conway2019}.
An exception to this rule is the bound state in the continuum (BIC), a localized state whose energy resides within the continuum spectrum. 
BICs in Hermitian systems are structurally fragile and typically require fine-tuning of system parameters. 
This fragility arises from the necessity for BICs to decouple from surrounding extended scattering states to remain localized. 
Such decoupling often relies on symmetry protection or interference-based mechanisms. For example, an odd-parity BIC can remain localized if the continuum states have even parity, preventing hybridization due to symmetry mismatch~\cite{Hsu2016}. 
Moreover, BICs in Hermitian systems exhibit exponential localization in arbitrary space dimensions. 
These limitations and characteristics motivate us to investigate BICs in non-Hermitian systems~\cite{Rotter_2009,Diehl2011,Simon2015,Jean1992,Regensburger2012,Gao2015_Nature,FengLiang2017,Ganainy2018,Miri2019,YangLan2019,FuLiang2017_arXiv,ShenHT2018_PRL,FuLiang2020_PRL,SongF2019_PRL,Ashida2020} through the following questions:
(i)~Is there a universal mechanism for the emergence of BICs in non-Hermitian systems that does not rely on symmetry protection or fine-tuning?
(ii)~What is the localization behavior for such BICs in non-Hermitian systems?

The questions above are both reasonable and compelling, given the fundamental differences in two key aspects between Hermitian and non-Hermitian systems: spectral structure and wavefunction localization.
Unlike Hermitian systems, where the energy spectrum lies along the real-energy axis, non-Hermitian systems typically feature complex energy eigenvalues that can span a finite area in the complex plane~\cite{Kai2022NC}. 
This results in a fundamentally distinct spectral structure and density of states, beyond the constraints of Hermitian systems. 
Moreover, non-Hermitian systems often exhibit the non-Hermitian skin effect(NHSE), where a macroscopic number of eigenstates become localized at system's boundaries~\cite{Lee2016PRL,luisPRB2018,Yao2018,Kunst2018_PRL,WangZhong2018,Murakami2019_PRL,ChingHua2019,LeeCH2019_PRL,LonghiPRR2019,Kai2020,Okuma2020_PRL,Slager2020PRL,Zhesen2020_aGBZ,Zhesen2020_SE,XuePeng2020,Ghatak2020,Thomale2020,LiLH2020_NC,Kawabata2020_Symplectic,Wanjura2020_NC,XueWT2021_PRB,LiLH2021_NC,Kai2022NC,ZhangDDS2022,Longhi2022PRL,YMHu2022}. 
Importantly, the non-Hermitian skin effect can exhibit either exponential or algebraic localization, depending on spatial dimensionality and other factors~\cite{Yao2018,LiLH2020_NC,Kawabata2023PRX,Zhang2025}.  
Although the non-Hermitian skin effect typically requires fully open boundaries, a single impurity that weakly breaks translational invariance can still exhibit localization behavior related to the skin effect. 
These key factors lead to the emergence of algebraic states in continuum (AICs) in non-Hermitian systems, which are fundamentally distinct from BICs in Hermitian systems. 

We first analytically investigate the emergence of AICs in a two-dimensional continuum model with a delta-function impurity potential. 
Our analysis shows that the appearance of AICs requires only that the system's Bloch spectrum spans a finite area in the complex-energy plane, indicating that AICs are a generic feature of non-Hermitian systems, without the need for fine-tuning. 
We further demonstrate that AICs can arise in periodic-boundary lattice systems in the thermodynamic limit and derive the threshold condition for the impurity strength required to induce them. 
Beyond these isolated AIC states, we find that the continuum states are modified into a coherent superposition of plane waves and AIC components. 
Additionally, a pronounced peak in the local density of states(LDOS) appears at the AIC energy, exhibiting a resonance-like feature reminiscent of Hermitian systems.

Here we emphasize how AICs differ from other localization phenomena. In Hermitian 2D media, a point scatterer generates an outgoing cylindrical wave with envelope $|\psi_{\mathrm{sc}}|\sim  r^{-1/2}$; this follows from flux conservation (constant $2\pi r\,j_r$) or, equivalently, from the Hankel/stationary-phase asymptotics of integrating over the one-dimensional constant-energy contour~\cite{Economou2006}. By contrast, in our non-Hermitian setting where the Bloch spectrum covers a finite area in the complex-energy plane, a single impurity yields AICs whose envelope is algebraically localized, with the form of $|\psi|\sim r^{-1}$. The change of exponent is tied to $k$-space geometry: the non-Hermiticity reduces the constant-energy manifold from a 1D curve to isolated points.

Moreover, while both NHSE and AICs depend on the same bulk criterion (a finite PBC spectral area)~\cite{Kai2022NC}, AICs are non-topological in nature.  
AICs require neither point-gap topology nor dislocations, remain $O(1)$ in number, and display the characteristic $r^{-1}$ profile. Thus AICs are fundamentally different from other topologically originated localization phenomena in non-Hermitian systems. For example, the dislocation induced non-Hermitian skin effect(DNHSE)~\cite{Schindler2021,Bhargava2021,Panigrahi2022,Xue2021,Wu2025}, 
which is a topological response of point-gapped bands, is triggered when the Burgers-vector line $B \cdot k=\pi$ winds nontrivially, and produces $O(L)$ exponentially localized modes at the dislocation core(s).

\emph{{\clr Algebraic states in continuum with a delta-type potential}.---}~Consider a two-dimensional non-Hermitian Hamiltonian $H_0(-i\partial_x,-i\partial_y)$ with a delta-function impurity potential $V=\lambda \, \delta(x,y)$, where $\lambda \in \mathbb{C}$ in general. 
The eigenvalues form a continuous spectrum in the thermodynamic limit.
In Hermitian systems, a given energy $E_0$ inside this continuum
typically corresponds to a continuum of momentum solutions: the
momentum-space solution set, 
\begin{equation}
	\begin{split}
		S(E_0) := \{\mathbf{p} | E_0 = H_0(\mathbf{p}),\mathbf{p}=(p_x,p_y)\in BZ \},  \label{eq:discrete_set}
	\end{split}
\end{equation}
forms a continuous contour in BZ. In contrast, for non-Hermitian systems whose spectrum spans a finite area in the complex-energy plane, $S(E_0)$ is discrete~\cite{Kai2022NC}. 
This difference in $k$-space structure underlies the emergence of AICs.

For simplicity, we assume that $E_0=H_0(\mathbf{p}_0)$ is nondegenerate and take $\mathbf{p}_0=\mathbf{0}$. A first-order expansion $H_0(\mathbf{p}) \approx E_0 + (\partial_{p_x} H_0) p_x + (\partial_{p_y} H_0) p_y$ reduces the eigenvalue equation to
\begin{equation}
	[-i\partial_x  -ia \partial_y + \tilde{\lambda} \delta(x,y)]\psi(x,y) = 0, \label{eq:dirac}
\end{equation}
where $\tilde{\lambda}= \lambda/\partial_{p_x} H_0|_{\mathbf{p}=\mathbf{p}_0}$, 
 and $a = \partial_{p_y} H_0/\partial_{p_x} H_0|_{\mathbf{p}=\mathbf{p}_0}$ is a non-real number. 
The fact that $\Im a\neq 0$ means that the system's spectrum around $E_0$ locally occupies a finite area and vice versa.
Eq.~\eqref{eq:dirac} can be solved via Fourier transform, yielding
\begin{equation}
	\psi(x,y) =\frac{\tilde{\lambda} \, \psi(0,0)}{2\pi i} \int dk_x dk_y   \frac{e^{i k_x x + i k_y y}}{k_x + a k_y } . \label{eq:dirac_fourier}	
\end{equation}
Without loss of generality, we assume $\Im(a)>0 , x>0 ,y >0$. Evaluating Eq.~\eqref{eq:dirac_fourier} by residue theorem~\cite{SupMat},we obtain
\begin{equation}
	\begin{split}
		\psi(x,y) = \tilde{\lambda} \, \psi(0,0) \frac{1}{y-ax} \propto \frac{c(\theta)}{r}   \label{eq:algebraic}
	\end{split}
\end{equation}
Notably, $\Im{a}\neq 0$ is crucial for the algebraic localization, which is guaranteed by the finite-area condition for continuum spectrum, thus making AICs unique to non-Hermitian system in two and higher dimensions. 
In the Hermitian case, where  $a \in \mathbb{R}$, the integral in Eq.~\eqref{eq:dirac_fourier} diverges, indicating the absence of such AIC solutions. This also aligns with the established theorem that BICs cannot arise in Hermitian systems with spatially confined impurity potentials~\cite{vonNeumann1931, Hsu2016}.  Notably, as the system transitions from Hermitian to weak non-Hermitian, AICs emerge once the spectrum acquires any finite area ,and $c(\theta)$ in Eq.~\eqref{eq:algebraic} is controlled by Jacobian of $H_0$ at $\mathbf{p}_0$,i.e., $\max_\theta |c(\theta)| \propto J(\mathbf{p}_0)^{-1}$~\cite{SupMat}.

 Note that, while we focus on delta-type potentials in two dimensions here, algebraic decay is in fact a generic feature of non-Hermitian systems in dimensions $d>1$. It arises from two ingredients: 
(i) the continuum spectrum occupies a finite region in the complex-energy plane, and
(ii) breaking translational invariance. 
Both conditions remain valid in higher dimensions and for more realistic
impurity potentials, provided that their tails decay sufficiently fast
(e.g., screened Coulomb or Gaussian potentials). 
Dimensionality, however, changes the decay exponent. 
In general $d>1$ dimensional case, AICs generally decay as $r^{-\frac{d}{2}}$ with $r^{-1}$ as exception for specific directions.
We provide a detailed derivation concerning above points in the Supplemental Material ~\cite{SupMat}. 

\emph{{\clr Lattice implementation and Green's function}.---}~
The algebraically decaying state described above can be explicitly established in a lattice model. 
We consider a two-dimensional non-Hermitian tight-binding Hamiltonian $H_0$ with PBC, perturbed by a single on-site impurity at the origin, $\lambda \ket{\mathbf{0}}\bra{\mathbf{0}}$ with $\lambda \in \mathbb{C}$.  
The full Hamiltonian is
\begin{equation}
	\begin{split}
		\hat{H} = \hat{H}_0+ \lambda \ket{\mathbf{0}}\bra{\mathbf{0}} ,\lambda \in \mathbb{C} . \label{eq:system_hamiltonian}
	\end{split}
\end{equation}

Using the Green's function method ~\cite{Fang2023,Yang2024,SupMat}, the eigenstate $\ket{\psi}$ with eigenvalue $E_0$ can be related to the Green's function $\hat{G}_0(E) = (E-\hat{H}_0)^{-1}$ as 
\begin{equation}
	\begin{split}
		\bra{\mathbf{x}}\ket{\psi} &= \lambda  \bra{\mathbf{x}}(E_0-\hat{H}_0)^{-1}\ket{\mathbf{0}}\bra{\mathbf{0}}\ket{\psi} . \label{eq:spatial_profile}
	\end{split}
\end{equation}
Inserting the resolution $\mathbb{I}=\sum_{\mathbf{k}\in BZ} \ket{\mathbf{k}}\bra{\mathbf{k}}$ and denoting the Bloch Hamiltonian by $H_0(\mathbf{k}) = \bra{\mathbf{k}}\hat{H}_0\ket{\mathbf{k}},\mathbf{k}=(k_x,k_y)$, we have
\begin{equation}
	\begin{split}
		G_0(E_0,\mathbf{x}):=\bra{\mathbf{x}}(E_0-\hat{H}_0)^{-1}\ket{\mathbf{0}}=  \sum_{\mathbf{k}\in BZ} \frac{\bra{\mathbf{x}}\ket{\mathbf{k}}\bra{\mathbf{k}}\ket{\mathbf{0}}}{E_0-H_0(\mathbf{k})} . \label{eq:Green}
	\end{split}
\end{equation}
Since the spatial profile $\psi(\mathbf{x})/(\lambda \psi(\mathbf{0}))$ corresponds directly to $G_0(E_0,\mathbf{x})$, we use these two quantities interchangeably.

\begin{figure}[t]
    \centering
    \includegraphics[width=1\linewidth]{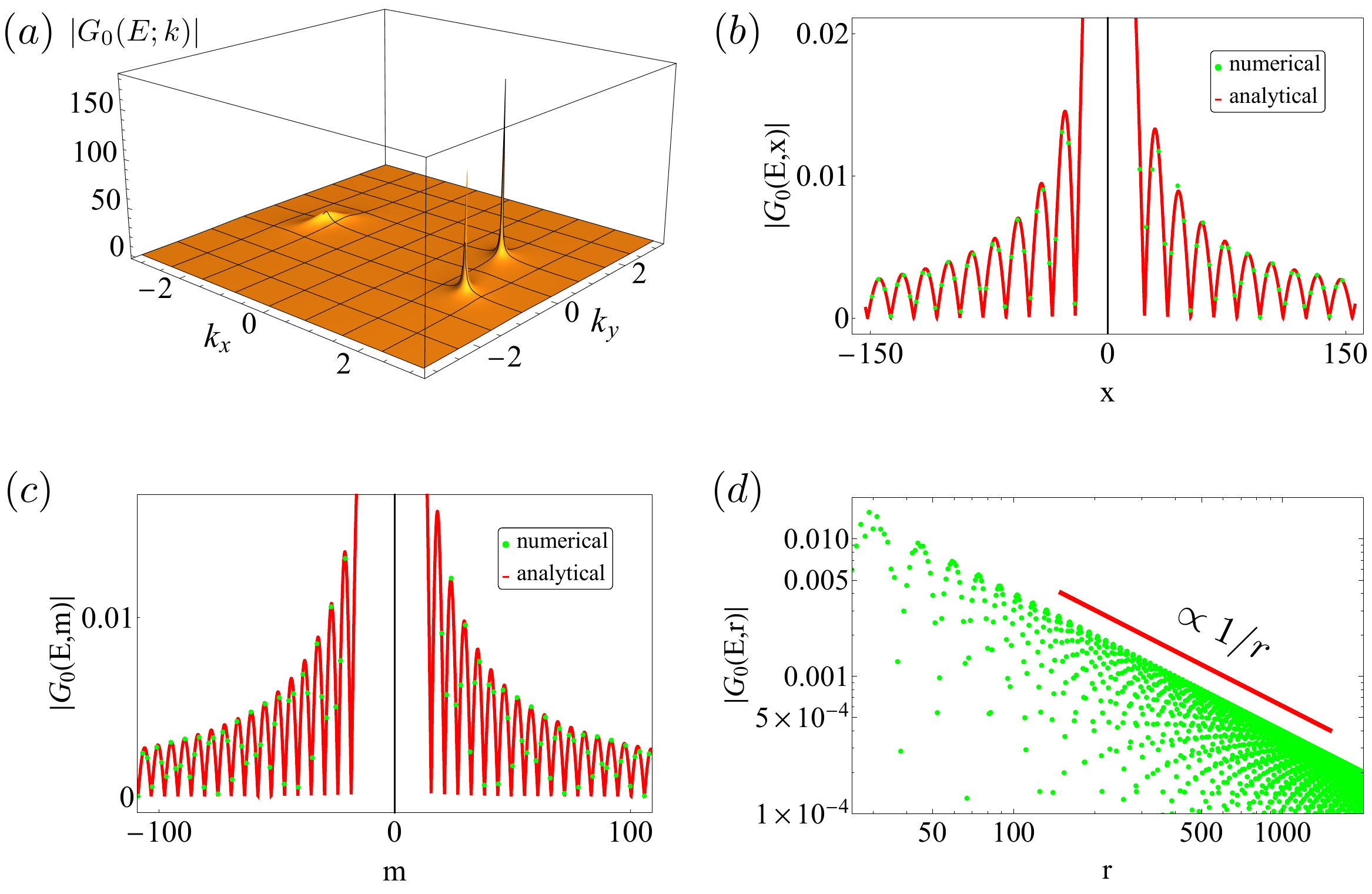}
    \caption{
	The AIC with $E_0=-0.21+0.98 i$ in model Hamiltonian $\tilde{H}_0(\mathbf{k}) = 0.6 e^{i k_x} + 0.4 e^{-i k_x } + 0.7 i e^{ik_y} + 0.4 ie^{-ik_y}$. (a) The absolute value of momentum-resolved Green's function, $|E_0 - \tilde{H}_0(\mathbf{k})|^{-1}$. 
	The two peaks represent two momentum solutions, $\mathbf{k}_0^{(1)} \approx (1.57, 0.78), \mathbf{k}_0^{(2)} \approx (2.00, -0.76)$. The spatial profile $G_0(E,\mathbf{x})$ along the x-axis in (b) and along the $x= y$ direction in (c). The green points are calculated from direct numerical integration of Eq.~\eqref{eq:Green}, and the red solid line is the asymptotic approximation calculated from Eq.~\eqref{eq:asymptotic} with the singularity points $\mathbf{k}_0^{(1)},\mathbf{k}_0^{(2)}$. (d)The log-log plot of the spatial profile along x axis.}
    \label{fig:f1}
\end{figure}

We first focus on AICs and their spatial profiles. In the thermodynamic limit, the summation over $\mathbf{k}$ in Eq.~\eqref{eq:Green} can be replaced by an integral~\footnote{The theory for bound state with energy outside the continuum spectrum has been established in Refs~\cite{Fang2023,Yang2024}.}. 
In contrast to the Hermitian case, where $S(E_0)$ forms a one-dimensional contour, and the associated singularity produces scattering states, in the non-Hermitian case the set $S(E_0)$ is discrete [Fig ~\ref{fig:f1}(a)] and no divergence occurs, as discussed above. 
These discrete momenta dominate the Green's function behavior at large $\lvert \mathbf{x}\rvert$. Thus,  Eq.~\eqref{eq:Green} can be approximated by expanding around these isolated solutions $\mathbf{k}_0\in S(E_0)$, 
\begin{equation}
	\begin{split}
		 \sum_{\mathbf{k}_0 \in S(E_0)} c(\mathbf{k}_0) e^{i \mathbf{k}_0 \cdot \mathbf{x}} \int d^2 \mathbf{k} \frac{e^{i \mathbf{k}\cdot \mathbf{x} }}{k_x + a(\mathbf{k}_0) k_y}, \label{eq:asymptotic}
	\end{split}
\end{equation}
where $c(\mathbf{k}_0) = - (2\pi)^{-2} (\partial_{k_x}H_0)^{-1}|_{\mathbf{k}_0}$ , and $a(\mathbf{k}_0) = \frac{\partial_{k_y}H_0}{\partial_{k_x}H_0}|_{\mathbf{k}_0}$. 
Each term in Eq.~\eqref{eq:asymptotic} reproduces the continuum solution for the delta impurity given by Eqs.~\eqref{eq:dirac} and~\eqref{eq:dirac_fourier}, so the eigenstate is a superposition of algebraically localized components, modulated by phase factors $e^{i \mathbf{k}_0 \cdot \mathbf{x}}$. 

A numerical illustration is shown in Fig.~\ref{fig:f1}. The model is  chosen as $\tilde{H}_0(\mathbf{k}) = 0.6 e^{i k_x}+ 0.4 e^{-i k_x } + 0.7 i e^{ik_y}+0.4 ie^{-ik_y}$ with energy $E_0 = -0.21 + 0.98 i$. 
Two discrete peaks in Fig.~\ref{fig:f1}(a) correspond to two singular points, $\mathbf{k}_0^{(1)} \approx (1.57, 0.78), \mathbf{k}_0^{(2)} \approx (2.00, -0.76)$, each giving rise to an algebraically localized component, as demonstrated in Figs.~\ref{fig:f1}(b) and (c). 
Numerical simulations (green points) show excellent agreement with the analytic results derived in Eq.~\eqref{eq:asymptotic} (solid curves).  The algebraic decay of the profile can be verified in the log-log plot[Fig.~\ref{fig:f1}(d)].   We emphasize that for $|\mathbf{x}|\gg 0$, the spatial profile is dominated by the singularity points, and the contribution from other detailed features, for example, the finite local maximum around $\mathbf{k'}\approx (-1.6,0)$ in Fig.~\ref{fig:f1}(a), can be neglected .

In the Supplemental Material~\cite{SupMat}, we demonstrate that the same result can be derived through asymptotic analysis as $|\mathbf{x}|\to \infty$. Specifically, we present a concrete example in which the algebraic decay emerges naturally by applying standard asymptotic expansion techniques to the Green's function. 

\emph{{\clr Threshold eliminated by Bloch Saddle Point}.---}~Now that the existence of algebraic impurity states in the continuum has been established, let us investigate the impurity strength threshold required to induce such states. 

Evaluating Eq.~\eqref{eq:spatial_profile} at $\mathbf{x} = 0$, we define 
\begin{equation}
	\begin{split}
		f_\lambda(E) :=\lambda^{-1} - G_0(E,\mathbf{0}),   \label{eq:imp_condition}
	\end{split}
\end{equation}
 and denote $P(\lambda)=\{E| f_\lambda(E)=0\}$. 
It follows that an AIC with energy $E_0$ is excited by an impurity with strength $\lambda$ only when $E_0\in P(\lambda)$, and its spatial profile is given by Eq.~\eqref{eq:asymptotic}. Thus, the number of AICs excited is given by the cardinality of $P(\lambda)$, which is $O(1)$.
Furthermore, this requirement defines a function relating the impurity strength $\lambda$ and the energy $E$ of AIC, i.e., $\lambda(E) =G_0(E,\mathbf{0})^{-1}$, with $E$ in the continuum spectrum of $\hat{H}_0$. 

Since $G_0(E,\mathbf{0})$ is continuous for $E$ in the continuum band, $\lambda(E)$ is also continuous there~\footnote{As a side remark, $\lambda(E)$ is in fact non-analytic. Since the singularity set depends on $E$ non-analytically, $\lambda^{-1}(E)$ also exhibits non-analytic behavior~\cite{SupMat}.}. Consequently, the image set, $\{\lambda(E)| E= H_0(\mathbf{k}),\mathbf{k} \in BZ\}$, forms a connected region in the complex plane. Physically, for a given system, only an impurity with strength $\lambda$ inside the image set of $\lambda(E)$ can generate an AIC. 

Thus, if the image set of $\lambda(E)$ covers the origin, i.e., $\lambda(E_c) =  0$, or equivalently, the integral for $G_0(E_c,\mathbf{0})$ diverges, for some $E_c$, no threshold impurity strength is required. 
Otherwise, there exists a finite threshold strength, explicitly given by $\min_{\mathbf{k}} |\lambda(H_0(\mathbf{k}))|$,necessary to induce the algebraically localized states. 

As discussed above, the integral defining $G_0(E,\mathbf{0})$ is convergent if the gradient of $H_0(\mathbf{k})$, $\nabla H_0(\mathbf{k})$ ,is non-zero for all $\mathbf{k}$
~\footnote{Technically, what we've shown applies to a system whose gradient is nonzero. Thus, to complete the proof of this statement, one needs to prove the convergence of integral if the integrand near the singularity behaves like $\frac{1}{k_x^n + a k_y} , a\in \mathbb{C}\setminus \mathbb{R} , n\ge 2$. We omit the mixing term $k_x^{n-m} k_y^{m}$ here since one can extract the common $k_y$ as $k_y(a+ O(k))$ which is dominated by $k_y$ near the origin. To show the convergence, one just needs to apply the residue theorem to integrate out $k_x$ and gets $\int k_y (k_y)^\frac{1-n}{n}$ which is convergent.}. 
Conversely, Ref~\cite{Yang2024} demonstrates that the integral diverges at Bloch Saddle Point (BSP, $\mathbf{k}_c$ such that $\nabla H_0(\mathbf{k}_c)=0$) if BSP exists~\footnote{We remark that the reference~\cite{Yang2024} deals with bound states outside the spectrum, i.e., $\lambda(E)$ for $E\in \mathbb{C}\setminus \{H_0(\mathbf{k}),\mathbf{k}\in BZ\}$. By Liouville's theorem, this function takes the minimum value at the boundary. So BSP needs to be at the boundary of the spectrum to eliminate the threshold. Since we are now considering states in the continuum, this constraint can be relaxed.}. 

Physically, this implies that a finite impurity threshold is required to induce algebraically localized states if and only if the group velocity, $\nabla H_0(\mathbf{k})$, is nonzero everywhere in the Brillouin zone.

\emph{{\clr AIC as scattering wavefunction}.---}~
In the Hermitian case, resonance provides a route to obtain localized, but unstable, scattering states in the continuum spectrum~\cite{Philip2012}.  Similarly, in non-Hermitian systems, the general solution of the eigenvalue problem for the Hamiltonian in Eq.~\eqref{eq:system_hamiltonian} can be expressed as
\begin{equation}
	\begin{split}
		\ket{\psi_\mathbf{k}} = \ket{\mathbf{k}}+  (E-\hat{H}_0)^{-1} \hat{V}\ket{\psi_\mathbf{k}} ,\label{eq:formal_solution}
	\end{split}
\end{equation}
where $\ket{\mathbf{k}}$ is a plane wave with eigenvalue $E$. Here, we omit $\pm i0^+$ prescription in Green's function that distinguishes ingoing/outgoing limits in Hermitian scattering.
More precisely, in higher-dimensional non-Hermitian systems with area-type spectrum,  the elements of Green's function in position space are well-defined, i.e., $\bra{\mathbf{r}}(E-\hat{H}_0)^{-1}\ket{\mathbf{r}'}$ is finite and decays algebraically as $|\mathbf{r}-\mathbf{r}'|^{-1}$ for all $\mathbf{r},\mathbf{r}'$ and $E$ within continuum spectrum, as shown above.  Physically, $\pm i0^+$ can be understood as the on-shell constraint during the scattering process via the Sokhotski-Plemelj formula~\cite{SupMat,Economou2006}. In non-Hermitian case, this on-shell constraint no longer holds in the absence of energy conservation.

\begin{figure*}[t]
	\begin{centering}
		\includegraphics[width=1\linewidth]{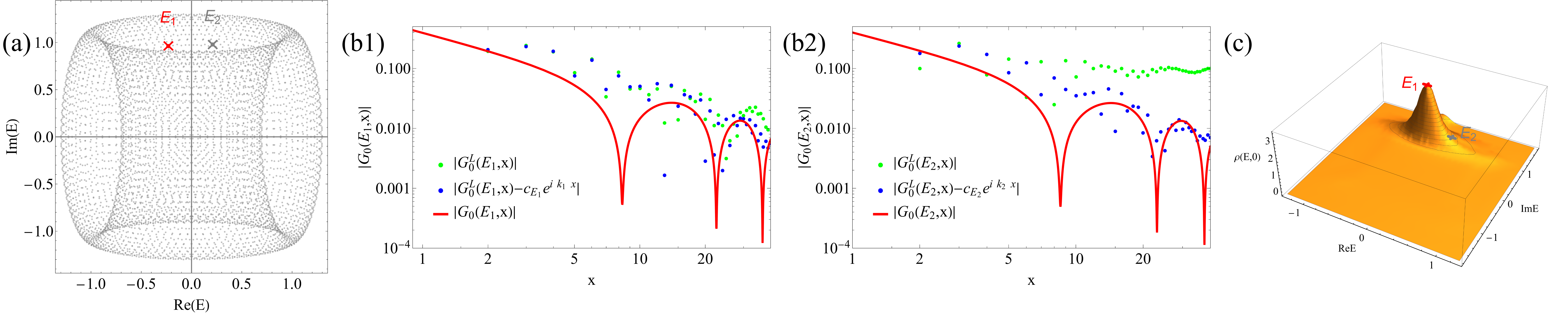}
		\par\end{centering}
		\protect\caption{
		(a) Perturbed spectrum of $\tilde{H}_0$ under a $80\times80$ lattice with impurity $\lambda_0$. The parameter is same as Fig.~\ref{fig:f1}. Red cross($E_1$) is the nearest eigenstate to $E_0$, and gray cross($E_2$) is an arbitrary state. Note that a bound state~\cite{Yang2024} with energy $E_b \approx -0.22+2.00i$ is not shown in the figure, since it's not our focus here.
		 (b1)[(b2)] Green dots: the spatial profile of $G^L_0(E_1,\mathbf{x})$[$G^L_0(E_2,\mathbf{x})$] along the $x$-axis; Blue dots: the spatial profile subtracted by the extended plane wave part; red solid line: the asymptotic approximation of the algebraic decay calculated from Eq.~\eqref{eq:asymptotic} . (c) The LDOS $\rho(E,\mathbf{0})$ of the system, with red cross and gray cross indicating the energy of $E_1$ and $E_2$ respectively.
		}\label{fig:f2}
\end{figure*}

For each plane wave $\ket{\mathbf{k}}$ , Eq.~\eqref{eq:formal_solution} defines a corresponding solution  $\ket{\psi_\mathbf{k}}$, thus it describes all continuum states of perturbed system $\hat{H}$~\cite{SupMat}. 
Furthermore, by specifying $\hat{V}= \lambda\ket{\mathbf{0}} \bra{\mathbf{0}}$ 
in Eq.~\eqref{eq:formal_solution} and evaluate it with $\bra{\mathbf{x}=\mathbf{0}}$,  one gets $\lambda \psi_\mathbf{k}(\mathbf{0}) f_\lambda(E) = \bra{\mathbf{0}}\ket{\mathbf{k}}$.  And thus Eq.~\eqref{eq:formal_solution} can be reduced into
\begin{equation}
	\begin{split}
		\ket{\psi_\mathbf{k}} = \lambda \psi_\mathbf{k}(\mathbf{0})  \left( f_\lambda(E)\frac{\ket{\mathbf{k}}}{\bra{\mathbf{0}}\ket{\mathbf{k}}} + \hat{G}_0(E)\ket{\mathbf{0}} \right). \label{eq:scattering_state}
	\end{split}
\end{equation}
Once recognize $\ket{AIC} =  \hat{G}_0(E)\ket{\mathbf{0}}$ as AIC part,
one can show that all states in the continuum spectrum can be written as a superposition of plane wave part and AIC part. And there are specific states with energy $E$ such that $f_\lambda(E)=0$,i.e,no plane wave part and fully algebraically localized. Those specific states are exactly AICs we discuss in previous sections. In supplemental material~\cite{SupMat}, we provide a more detailed discussion on finite lattice case.

In Fig.~\ref{fig:f2}, an illustration of the structure of perturbed bulk eigenstates is presented. Fig.~\ref{fig:f2}(a) shows the spectrum  of $\tilde{H}_0$ on an $80\times80$ lattice with impurity $\lambda_0 = G_0(E_0,\mathbf{0})^{-1} \approx -0.20+2.04 i$, corresponding to AIC at $E_0$ in the thermodynamic limit. 
We focus on the spatial profile of the eigenstate closest to $E_0$, labeled $E_1$(red cross), and another arbitrarily chosen state $E_2$(gray cross).
Since $E_1$ is close to $E_0$, its plane wave component is negligible, and the spatial profile is dominated by the algebraically localized component. In contrast, $E_2$ exhibits a significant plane wave component. Quantitatively, Figs.~\ref{fig:f2}(b1) and (b2) show the spatial profile of $E_1$ and $E_2$ along the x-axis (indicated by green dots), respectively. After the subtraction of the plane wave component(blue dots), the remaining term can be well-approximated by AIC calculated in Eq.~\ref{eq:asymptotic} (red solid line). 

\emph{{\clr LDOS peak at AIC energy}.---}~In Hermitian systems, a characteristic signature of a resonance is a peak in the local density of states (LDOS) at the impurity site~\cite{Economou2006}. The LDOS at the impurity $\mathbf{0}$ is defined as the density of states weighted by the wavefunction amplitude at the impurity, 
\begin{equation}
	\begin{split}
		\rho(E,\mathbf{0}) := \sum_i \delta(E-E_i) |\psi_i(\mathbf{0})|^2.  \label{eq:partial_DOS}
	\end{split}
\end{equation}
In non-Hermitian systems, the extension of this definition is not
unique, since one may choose the weight to be the amplitude of the left, right, or biorthogonal eigenstates. As we shall see, however, this choice does not affect the qualitative conclusions below. For
definiteness, we use the right-eigenstate weight $|\psi_i^R(\mathbf{0})|^2$. 

In non-Hermitian systems, there is also a peak at the AIC energy for $\rho(E,\mathbf{0})$. 
For generic continuum energies, the corresponding eigenstate is a superposition of a localized (AIC-like) component and a plane-wave component, with the latter vanishing exactly at the AIC energy. Since the plane-wave part generally reduces the weight at the impurity site by delocalizing the wavefunction, states with energies near the AIC energy carry larger weight than typical continuum states. Consequently LDOS develops a peak at the AIC energy.
As a demonstration, Fig.~\ref{fig:f2}(c) shows the LDOS at the impurity site for the model, with a pronounced peak located at $E_0$.

 Direct in situ measurement of the LDOS at a complex eigenenergy is indeed challenging, especially in quantum platforms, because standard local probes access the response only at real excitation frequencies. However, the LDOS and the AIC wavefunction can be inferred indirectly by reconstructing the Green's function from real-frequency measurements. This type of Green's-function tomography has already been implemented in higher-dimensional acoustic lattices, where complex spectra and biorthogonal eigenmodes are extracted directly from experiment ~\cite{Zhong2025}.

\emph{{\clr Discussions and conclusions}.---}~Since the number of AICs,  i.e., the cardinality of $P(\lambda)$, is of order one, they constitute only a negligible fraction of the system's total degrees of freedom. One might question their experimental accessibility. This issue can be addressed using a perturbative approach: for a generic continuum eigenstate, one can compare its spatial profile with and without the impurity.This difference is expected to exhibit algebraic decay in space, indicating a clear experimental signature. 

 As a final remark, although we focus on PBC throughout, the concept of AICs can be extended to OBC~\cite{SupMat}. Specifically, for systems with corner skin effect ~\cite{Wang2024,Zhang2025,Zhang2024_edge,Xu2023,Xiong2024}, algebraically modified states also exist in the sense that they correspond to solutions of $f_\lambda(E)=0$ within continuum spectrum. Intuitively, with the exponential suppression of skin effect, such states generally behave as $O(|\mathbf{r}-\mathbf{r}_{\text{imp}}|^{-1}e^{-\boldsymbol{\mu}\cdot\mathbf{r}})$ with $\mathbf{r}_{\text{imp}}$ as the position of the impurity. For systems with algebraic skin effect ~\cite{Zhang2025},the generalization of AICs is subtle since bi-orthogonal decomposition is challenging in this case. We conjecture that AICs are still greatly suppressed and hard to observe as suggested by numerics~\cite{SupMat}, given that skin modes are rather robust against local perturbation.

We emphasize that in non-Hermitian systems where the skin effect is absent under specific open boundary geometries --- for example, systems exhibiting geometry-dependent skin effects (GDSE)~\cite{Kai2022NC,Wang2022NC,QYZhou2023NC,DingKun2023PRL,WanTuoSciB,QinYi2023PRA,Zhao2025} --- the AICs behave analogously to the PBC case and thus remain observable. Moreover, this class of systems is known to generically host Bloch saddle points~\cite{Yang2024}, implying that even infinitesimal impurity potential is sufficient to induce algebraically localized states embedded in the continuum spectrum.  

\section{Acknowledgements}
We thank Zixi Fang for helpful discussions.
C.F. acknowledges funding support by
National Natural Science Foundation of China (NSFC) under Grant Nos. 12325404, 12547112, and 12188101,
and National Key R\&D Program of China under Grant Nos. 2022YFA1403800 and 2023YFA1406704.

\bibliographystyle{apsrev4-2}
\bibliography{Refs_MainText}

\newpage 
\appendix
\setcounter{equation}{0}  %  this will re-count eq from 1
\setcounter{figure}{0}  %  this will re-count figure from 1
\renewcommand{\thefigure}{A\arabic{figure}}
\renewcommand{\theequation}{A\arabic{equation}}
\begin{widetext}
{\begin{center}
		{\bf \large Supplemental Material:Algebraic States in Continuum in d\textgreater1 Dimensional Non-Hermitian Systems}
\end{center}}

\section{Appendix I:Derivation of Eq.~(4) in the main text}
We start by arguing that the integration 
\begin{equation}
    \begin{split}
    \psi(x,y) =\frac{\lambda \psi(x=0,y=0)}{2\pi i} \int dk_x dk_y   \frac{e^{i k_x x + i k_y y}}{k_x + a k_y } . \label{eq:dirac_fourier}	
    \end{split}
\end{equation}
is convergent for $a \notin \mathbb{R}$. To see this, we transform the integral into the polar variables $\{k_r,k_\theta\}$. The absolute value of the integrand now scales as $1/k_r$, but it's multiplied by Jacobian $k_r$, ensuring that the singularity near $k_x=k_y=0$ does not bring out divergence. As for the divergence from infinity of the integral bound, it can be avoided with a suitable cutoff.

To show that the solution Eq.~\ref{eq:dirac_fourier} is algebraically localized, we evaluate it by applying the residue theorem to one variable, say $k_x$, and then integrate over the remaining variable.
Let's assume $\Im(a)>0 , x>0 ,y >0$ so that the integral contour of $k_x$ can be closed in the upper half plane.
Note that when  $k_y > 0$ , the integral over  $k_x$  vanishes because the relevant pole  $k_x = -a k_y$  lies in the lower half-plane. 
The integral over $k_y$ then reduces to 
\begin{equation}
	\begin{split}
		\int_{-\infty}^0 \text{exp}(i(y-ax)k_y) dk_y,
	\end{split}
\end{equation}
which exhibits algebraic decay behavior as $1/(y-ax)\propto 1/r$. 

\section{Appendix II: A concrete example of AIC}
In this section, we explicitly demonstrate the existence and property of the algebraic impurity states with a concrete example, namely,
\begin{equation}
    \begin{split}
        H&= H_0 + V, \\
        h_0(k_x,k_y) &= \cos k_x + i \cos k_y ,\\
        V &= \lambda \ket{0}\bra{0}  .
    \end{split}
\end{equation}
We begin by briefly revisiting Green's function formulation for impurity-induced states in non-Hermitian systems. Next, we analyze the dependence $\lambda(E)$ for this model. Interestingly, $\lambda(E)$ exhibits non-analytic behavior in the thermodynamic limit, which contrasts sharply with the finite case. Finally, we explicitly derive the asymptotic behavior of the algebraic impurity states in this mode.

\subsection{Revisit of Green function formulation for impurity induced states for general non-Hermitian systems}
Consider a general tight-binding model with a non-Hermitian Hamiltonian $\hat{H}_0$,
\begin{equation}
    \begin{split}
        \hat{H}_0 &= \sum_{\bm{r}}\sum_{\bm{s}} t_{\bm{s}} \ket{\bm{r}}\bra{\bm{r+s}} \\ &= \sum_{\bm{k}} \sum_{\bm{s}}  t_{\bm{s}}e^{i \bm{k\cdot s} }\ket{\bm{k}}\bra{\bm{k}}. \label{eq:hamiltonian}
    \end{split}
\end{equation}
Here we denote the Bloch Hamiltonian as $h_0(\bm{k})=\sum_{\bm{s}} t_{\bm{s}}e^{i \bm{k\cdot s}}$ , where $\bm{r}$ represents lattice site, $\bm{s}$ denotes the direction of hopping and $t_{\bm{s}} \in \mathbb{C} \setminus 0 $ is the corresponding hopping strength. The number of terms in $h_0(\bm{k})$ is finite since we consider a finite range of hopping. 

To solve the Schrödinger equation for the system with an impurity potential $V=\lambda \ket{\bm{0}}\bra{\bm{0}}$,
\begin{equation}
    (\hat{H}_0 + \lambda \ket{\bm{0}}\bra{\bm{0}}) \ket{\psi} = E \ket{\psi},
\end{equation}
Move the term $\hat{H}_0 \ket{\psi}$ to the right-hand side and take the inverse of $(E-\hat{H}_0)$. And the impurity states are determined by the Green's function of the system, which can be expressed as
\begin{equation}
    \begin{split}
        \frac{\psi(\bm{r})}{\lambda \psi(\bm{0})} = \bra{\bm{r}}(E-\hat{H}_0)^{-1}\ket{\bm{0}} = \frac{1}{(2\pi)^d} \int_{\bm{k}\in BZ} d^d \bm{k} \frac{e^{i \bm{k\cdot r}}}{E-h_0(\bm{k})}. \label{eq:Green}
    \end{split}
\end{equation} 
Here $d$ is the system's dimension. Unless stated otherwise, we focus on case $d=2$ .

\subsection{Non-analyticality of the Green function}
We now analyze the dependence of $\lambda(E)$ for the model $h_0(k_x,k_y) = \cos k_x + i \cos k_y$, which is given by integral
\begin{equation}
    \begin{split}
        \lambda^{-1}(E) = \frac{1}{(2\pi)^2} \int_{-\pi}^{\pi} dk_x \int_{-\pi}^{\pi} dk_y \frac{1}{E-\cos k_x - i \cos k_y}.
    \end{split}
\end{equation}
Apply the residue theorem to one of the variables, say $k_y$, and denote the roots for the equation $E-\cos k_x -i \frac{z+z^{-1}}{2}=0$ as $z_{1,2}(k_x,E)=-i(E-\cos k_x)\pm i\sqrt{(E-\cos k_x)^2 +1 }$, we have
\begin{equation}
    \begin{split} \label{eq:lambdaE}
        \lambda^{-1}(E) = \frac{i}{2\pi} \int_{-\pi}^{\pi} dk_x \frac{\chi_{|z_1(k_x,E)|<1}}{z_1(k_x,E) - z_2(k_x,E)} +   \frac{\chi_{|z_2(k_x,E)|<1}}{z_2(k_x,E) - z_1(k_x,E)} . 
    \end{split}
\end{equation}
Here,$\chi_A(k_x)$ is the characteristic function, i.e.$\chi_A(k_x) = 1 $ if $k_x\in A$ and $\chi_A(k_x)=0$ otherwise.  This function comes from the fact that we choose the pole inside the unit circle. Since $E$ is inside the continuum spectrum, i.e., $Re(E),Im(E) \in (-1,1)$ (the energy at the edge of the spectrum can be treated similarly with some modifications), there are four critical points for $E$, which we denote as $(\pm k_{x0}, \pm k_{y0}), k_{x0},k_{y0}\in (0,\pi)$. Thus, the integral Eq.~\eqref{eq:lambdaE} can be divided as 
\begin{equation}
    \begin{split}
        \lambda^{-1}(E) = \frac{1}{2\pi} (\int_{-k_{x0}}^{k_{x0}} - \int_{k_{x0}}^{2\pi-k_{x0}}) dk_x \frac{1}{\sqrt{(E - \cos k_x)^2 + 1}}. \label{eq:lambdaE2}
    \end{split}
\end{equation}
To illustrate the discussion above, let's take $E=\frac{1}{2} + \frac{\sqrt{2}}{2} i$ so that the four critical points are $(\pm \frac{\pi}{3},\pm \frac{\pi}{4})$. In Fig~\ref{fig:sm1}(a) we plot the function $z_{1,2}(k_x)$ in the range $k_x\in [0,\pi)$ in the complex plane. Note that $z_{1,2}(k_x,E)$ cross the unit circle at $k_x =\pm k_{x0}$, and the cross points are $e^{\pm i k_{y0}}$.
\begin{figure}[h]
    \centering
    \includegraphics[width=0.7\linewidth]{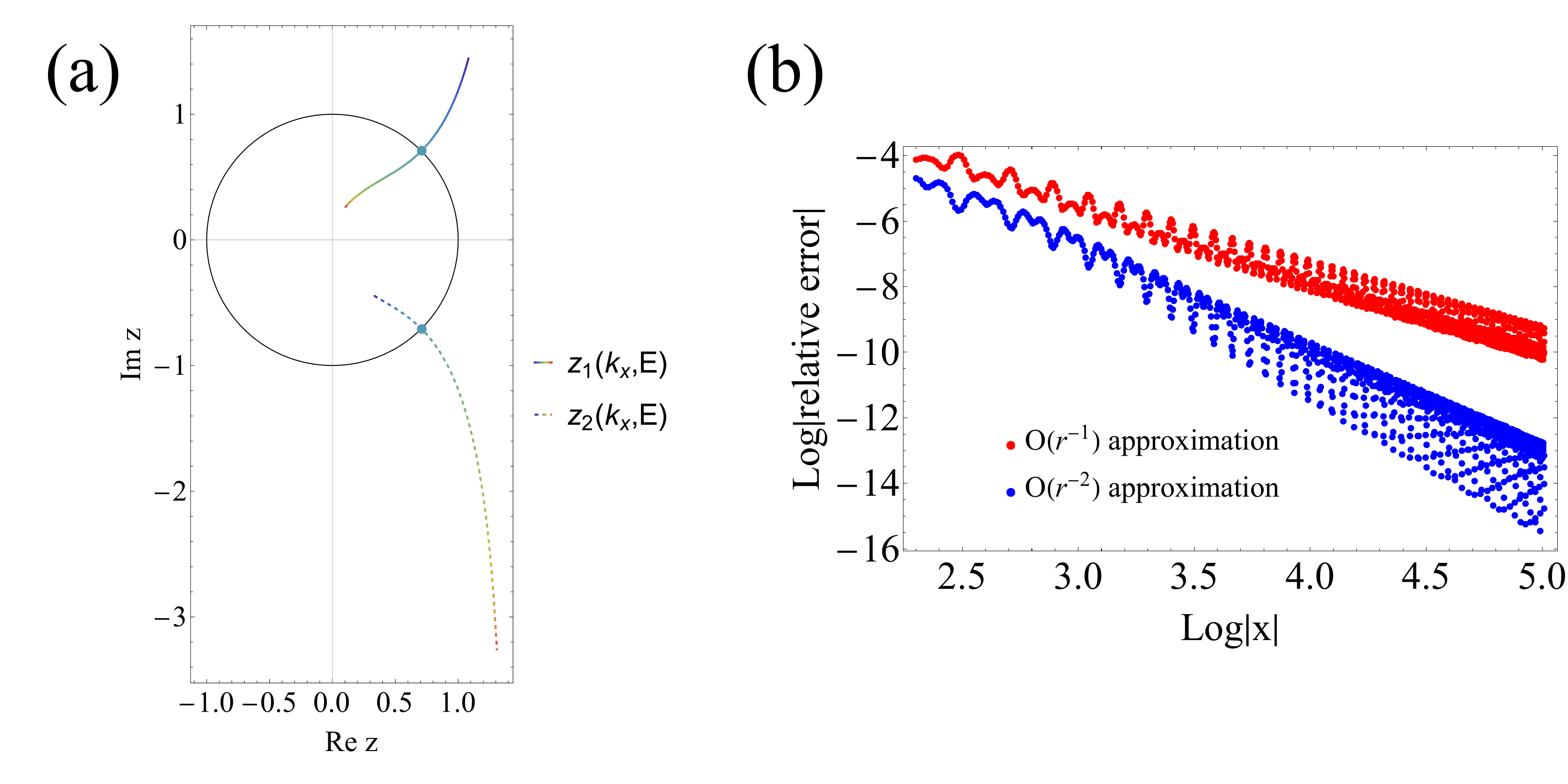}
    \caption{(a)The plot of $z_{1,2}(k_x,E)$ in the complex plane with $E=\frac{1}{2} + \frac{\sqrt{2}}{2} i$ and $k_x \in [0,\pi)$. The other part,$k_x\in [\pi,2\pi)$,is just ``backward" of the trajectory. Since we have $k_{x_0}= \frac{\pi}{3}$ and $k_{y_0}=\frac{\pi}{4}$, $z_{1,2}(k_x,E)$ cross the unit circle at $k_x = k_{x0}$ and the corresponding cross points are $e^{\pm i k_{y_0}}$
    (b) The absolute value of relative error between the exact numerical integration of Eq.~\eqref{eq:wavefunction} and asymptotic approximation Eq.~\eqref{eq:wavefunction2} truncated at $r^{-1}$ order (red points) and $r^{-2}$ order (blue points). 
    }
    \label{fig:sm1}
\end{figure}

Use the standard substitution $\tan \frac{k_x}{2} \to t_x$ and take advantage of the symmetry of the integrand, Eq.~\eqref{eq:lambdaE2} can be simplified into the form of the incomplete elliptic integral of the first kind, 
\begin{equation}
    \begin{split}
        \lambda^{-1}(E) = \frac{2}{\pi\sqrt{(E-1)^2 +1}} (\int_0^{t_{x0}}-\int_{t_{x0}}^\infty) dt_x \frac{1}{\sqrt{1-\frac{1-i+E}{1+i-E}t_x^2}\sqrt{1-\frac{1+i+E}{1-i-E}t_x^2}}.\label{eq:lambdaE3}
    \end{split}
\end{equation}
If there was no minus sign in Eq.\eqref{eq:lambdaE3}, the integral would be a complete elliptic integral of the first kind, which is precisely the case for $E$ outside the continuum spectrum\cite{Yang2024}. However, the minus sign in Eq.~\eqref{eq:lambdaE3} makes the integral non-analytic. To see this, note that $t_{x0}$, or equivalently $k_{x0}$, is determined by $E$ non-analytically,i.e. , $\cos k_{x0} = \Re E = \frac{E+E^*}{2}$. This non-analyticity is reflected in the integral Eq.~\eqref{eq:lambdaE3} as well.

Such non-analyticality is general for the Green's function of non-Hermitian systems for energy inside the spectrum. Indeed, from Eq.~\eqref{eq:Green}, we can see that the Green's function is related to the poles of the integrand. Such poles are non-analytic with respect to $E$, since they are determined by $h(\bm{z})=0, \{|z_1|,..,|z_d|\}=\{1,..,1\}$ and one immediately sees that the second relation is non-analytic.

\subsection{Asymptotic behavior of AIC}
Following derivation above, the wavefunction can be expressed as
\begin{equation}
    \begin{split}
        \frac{\psi(m,n)}{\lambda \psi(0,0)} &=  \frac{1}{(2\pi)^2} \int_{-\pi}^{\pi} dk_x \int_{-\pi}^{\pi} dk_y \frac{e^{i (k_x m + i k_y n)}}{E-\cos k_x - i \cos k_y} \\
        &= \frac{-i}{\pi} (\int_{-k_{x0}}^{k_{x0}} \frac{e^{i m k_x + n \ln z_1(k_x,E)}}{\sqrt{(E-\cos k_x)^2 +1}} - \int_{k_{x0}}^{2\pi-k_{x0}} dk_x \frac{e^{i m k_x + n \ln z_2(k_x,E)}}{\sqrt{(E-\cos k_x)^2 +1}}).
    \end{split} \label{eq:wavefunction}
\end{equation}
Use polar coordinate and denote $r(\cos \theta, \sin \theta):= (m,n)$, one can show that the wavefunction Eq.~\eqref{eq:wavefunction} asymptotically behaves as $\frac{\alpha(\theta,E)}{r}$, i.e., decays algebraically with direction-dependent coefficient. To see this, let's evaluate the first part of Eq.~\eqref{eq:wavefunction} at the x-axis, i.e.,$\theta=0,(m,n)=(r,0)$ as an example. After iterated integration by parts, it becomes (for simplicity, let's define $f(k_x):= 1/\sqrt{(E-\cos k_x)^2 +1}$)
\begin{equation}
    \begin{split}
        \int_{-k_{x0}}^{k_{x0}} dk_x f(k_x)e^{ik_x r} &= \frac{1}{ir} f(k_x)e^{ik_x r}|^{k_x = k_{x0}}_{k_x=-k_{x0}}  + \frac{-1}{ir} \int_{-k_{x0}}^{k_{x0}} dk_x  f'(k_x) e^{ik_x r} \\
        &= e^{ik_x r}(\frac{f(k_x)}{ir}-\frac{f'(k_x)}{(ir)^2}+...+(-1)^{n-1} \frac{f^{(n-1)}(k_x)}{(ir)^n})|^{k_x = k_{x0}}_{k_x=-k_{x0}} +\frac{(-1)^n}{(ir)^n} \int_{-k_{x0}}^{k_{x0}} dk_x  f^{(n)}(k_x) e^{ik_x r}  \label{eq:wavefunction2}
    \end{split}
\end{equation}
It follows from Riemann-Lebesgue Lemma that the remaining term $\frac{1}{r^n}\int_{-k_{x0}}^{k_{x0}} dk_x  f^{(n)}(k_x) e^{ik_x r} \sim o(r^{-n})$  as $r\to \infty$ \cite{Miller2006sub}. The second part of integral in Eq.~\eqref{eq:wavefunction} can be treated similarly, and we can now conclude that the wavefunction decays as $1/r$ in the x-axis. 

Note that the same argument can be applied to other directions and implies that the wavefunction decays as $\frac{\alpha(\theta,E)}{r}$ with $\alpha(\theta,E)$ direction-dependent coefficient.

However, some technical details need to be addressed to make the argument for arbitrary direction rigorous. One may suspect that the integral in Eq.~\eqref{eq:wavefunction}, after suitable deformation of contours, should be dominated by the saddle point of the exponent and thus decays exponentially, which follows from the steepest descent method. This is not the case, since one needs to keep the endpoints of the integral in Eq.~\eqref{eq:wavefunction} fixed while deforming the contour. The contribution from endpoints is $O(1/r)$, and  it's asymptotically larger than the contribution from the saddle point, which scales as $O(e^{-\alpha r})$.  

In Fig~\ref{fig:sm1}(b), we present the relative error between the exact numerical integration of Eq.~\eqref{eq:wavefunction} and the asymptotic approximation Eq.~\eqref{eq:wavefunction2}. As implied by Eq.~\eqref{eq:wavefunction2}, if one truncates the series at $r^{-1}$ order, the relative error is $O(1/r^2)$(red points), and if one truncates the series at $r^{-2}$ order, the relative error is $O(1/r^3)$(blue points). In principle, one can get arbitrary precision by truncating the series at higher order once the Hamiltonian and energy are known.

\section{Appendix III: Asymptotic Expansion of general Green's function}
The asymptotic method applied above can be generalized to arbitrary non-Hermitian systems, and it can be shown that the leading term is just what we get in the main text. To see this, let's derive the asymptotic expansion of the Green's function,
\begin{equation}
    \begin{split}
        G(x,y,E) &= \int_{-\pi}^\pi \frac{dk_x}{2\pi} e^{i k_x x}\int_{z_y\in \mathbb{S}^1} \frac{dz_y}{2\pi i z_y} \frac{z_y^y}{E-h_0(k_x,z_y)},
    \end{split}
\end{equation}
where we abuse the notation of $h_0(k_x,z_y)$ to denote the Bloch Hamiltonian $h_0$ after the variable transformation $z_y = e^{i k_y}$.  Without loss of generality, we assume $x\gg 0$ and $y\gg 0$ so that we pick the root of $z_y$ inside the unit circle when applying the residue theorem to $z_y$. Denote the roots as $z_{i}(k_x,E),i=1,...,n$, and we have 
\begin{equation}
    \begin{split}
        G(x,y,E) = \sum_{i=1}^n \int_{-\pi}^\pi \frac{dk_x}{2\pi} e^{i k_x x} z_i(k_x,E)^y  g_i(k_x)\chi_{|z_i(k_x,E)|<1} , \label{eq:Green2}
    \end{split}
\end{equation}
where we define $g_i(k_x) = Res_{z_y = z_i} \frac{1}{E-h_0(k_x,z_y)}$, and we assume $y-1\approx y$ since $y\gg 0$. For simplicity, we assume the term $\chi_{|z_i(k_x,E)|<1}$ restricts the integration region to $(k^i_l,k^i_u)$. Thus, Eq.~\eqref{eq:Green2} can be rewritten as
\begin{equation}
    \begin{split}
        G(x,y,E) = \sum_{i=1}^n \int_{k^i_l}^{k^i_u} \frac{dk_x}{2\pi} g_i(k_x)e^{irf_i(k_x)} , \label{eq:Green3}
    \end{split}
\end{equation}
where $x\equiv r \cos \theta , y \equiv r \sin \theta,  f_i(k_x) = k_x \cos \theta - i \sin \theta \ln z_i(k_x,E)$. Recall Eq.~\eqref{eq:wavefunction2} and related argument, we can see that the leading term is $O(1/r)$ and the related coefficient can be shown to be exactly the same as the one in the main text. 

\section{Appendix IV: Transition from Hermitian to non-Hermitian}
As the system transitions from Hermitian to non-Hermitian, with the spectrum shifting from purely real to occupying a region of nonzero area in the complex-energy plane, AICs appear once the spectrum has any finite area. Though the $r^{-1}$ tail is robust,the overall prefactor $c$ in $cr^{-1}$ is controlled by the width $W$(or more precisely, the local area $S$) of the spectrum in the narrow-strip limit.  \\

\textbf{(1)Transition from Hermitian to non-Hermitian.}\\

The crossover is most transparently seen by linearizing $H_0(\mathbf p)$ near a momentum $\mathbf p_0$ that solves $E_0 = H_0(\mathbf p_0)$. To model the transition, we start with a Hermitian system where the gradients $v_x = \partial_{p_x} H_0|_{\mathbf{p}_0}$ and $u = \text{Re} \partial_{p_y} H_0|_{\mathbf{p}_0}$ are both real. We then introduce non-Hermiticity via an imaginary component to the $p_y$ gradient, $i\eta = i \text{Im} \partial_{p_y} H_0|_{\mathbf{p}_0}$, i.e.,
\begin{equation}
    \begin{split}
        H_0(\mathbf p_0+\delta\mathbf p)\;\approx\;E_0\;+\;v_x\,\delta p_x\;+\;(u+i\eta)\,\delta p_y.
    \end{split}
\end{equation}

\textbf{Hermitian limit:$\eta=0$.}\\ 

The spectrum near $E_0$ traces a one-dimensional curve $E_0 + v_x\delta p_x + u\delta p_y\in\mathbb R$, the impurity integral diverges, and only extended scattering states exist.\\
    
\textbf{Turning on weak non-Hermiticity.}\\

For small but finite   $0<|\eta|\ll \min\{|u|,|v_x|\}$, the image of a small real rectangle in
$(\delta p_x,\delta p_y)$ around $\mathbf p_0$ becomes a narrow 2D strip in the complex-energy plane with local width
    \begin{equation}
        \begin{split}
            W \;\simeq\; 2\,|\eta|\,\Delta p_y,
        \end{split}
    \end{equation}
where $\Delta p_y$ is the range of $\delta p_y$ over which the linearization holds. In this regime the equation $E=H_0(\mathbf p_0+\delta\mathbf p)$ has only discrete solutions in the linearized patch, the Green-function integral converges, and   impurity-induced AICs exist. \\

The corresponding real-space wavefunction follows from Eq. (4) in the main text by residue evaluation:
\begin{equation}
    \begin{split}
        \psi(x,y)\;\sim \lambda \psi_0\frac{1}{(\partial_{p_x}H_0)y-(\partial_{p_y}H_0)x}  =\lambda\,\psi_0\,\frac{1}{v_x\,y-(u+i\eta)\,x} \;=\;\frac{\lambda\,\psi_0}{r}\,\frac{1}{v_x\sin\phi-u\cos\phi-i\eta\cos\phi}, \label{RB9}
    \end{split}
\end{equation}
where $x=r\cos\phi,\,y=r\sin\phi$ and $\psi_0= \psi(x=0,y=0)$.  One immediately sees that the
exponent is fixed while the prefactor tracks the strip width: the $1/r$ envelope is present for any $\eta\neq 0$. In the narrow-strip limit $0<|\eta|\ll |u|,|v_x|$,
\begin{equation}
    \begin{split}
   \big|\psi(r,\phi)\big|\;\approx\;\frac{|\lambda\psi_0|}{r}\,
   \frac{1}{\sqrt{\big(v_x\sin\phi-u\cos\phi\big)^2+(\eta\cos\phi)^2}} .
    \end{split}
\end{equation}
Along the ``most radiative'' direction $\phi_\ast$ defined by $v_x\sin\phi_\ast=u\cos\phi_\ast$, one finds
\begin{equation}
    \begin{split}
        \max_\phi |\psi(r,\phi)|\;\sim\;\frac{C}{r\,|\eta|} \propto \frac{1}{r\,W} \qquad(C=|\lambda\psi_0|/|\cos\phi_\ast|) ,
    \end{split}
\end{equation}
i.e. the $1/r$ prefactor scales as $1/W$. This controlled $1/W$ divergence as $\eta\to 0$ is precisely the way in which the non-Hermitian AIC solution approaches the divergence of the Hermitian impurity integral. \\

\textbf{(2)Generalization:Relationship between local area and algebraic prefactor.}\\

Treat the Bloch Hamiltonian $H_0(\mathbf p)$ as a general map,
\begin{equation}
    \begin{split}
        H_0:\mathbb{R}^2\to\mathbb{C}\cong\mathbb{R}^2,\quad \mathbf p=(p_x,p_y)\mapsto (E_r,E_i)=(\Re H_0(\mathbf p),\Im H_0(\mathbf p)).
    \end{split}
\end{equation}
Then the local area $S(\mathbf{p}_0)$ occupied by the spectrum near $\mathbf p_0$ is given by the Jacobian determinant,
\begin{equation}
    \begin{split}
        J(\mathbf{p}_0) = \det \frac{\partial(E_r,E_i)}{\partial(p_x,p_y)}\Big|_{\mathbf{p}=\mathbf{p}_0} = \Im (\partial_{p_x} H_0^* \cdot \partial_{p_y} H_0)\Big|_{\mathbf{p}=\mathbf{p}_0}, \label{RB13}
    \end{split}
\end{equation}
with $S(\mathbf{p}_0) = J(\mathbf{p}_0)\delta p_x \delta p_y$. \\

From Eq.~(4) in the maintext and Eq.~\eqref{RB9} above, the impurity-induced AIC wavefunction near $\mathbf p_0$ behaves as
\begin{equation}
    \begin{split}
        \psi(r,\phi)\sim \frac{\lambda \psi_0}{rf(\phi)} ,\quad f(\phi) = (\partial_{p_x}H_0)\sin\phi - (\partial_{p_y}H_0)\cos\phi.  \label{RB14}
    \end{split}
\end{equation}
Denote $||\nabla H_0||^2 = | (\partial_{p_x}H_0)|^2+ |(\partial_{p_y}H_0)|^2 $, and the mininum of $|f(\phi)|$ is given by 
\begin{equation}
    \begin{split}
        \min |f(\phi)|^2 &= \frac{||\nabla H_0||^2}{2}  - \frac{1}{2}\sqrt{||\nabla H_0||^4 - 4J(\mathbf{p}_0)^2} \\
        &\approx \frac{J(\mathbf{p}_0)^2}{||\nabla H_0(\mathbf{p}_0)||^2}  \label{RB15}
    \end{split}
\end{equation}
when $|J(\mathbf{p}_0)|\ll ||\nabla H_0(\mathbf{p}_0)||^2$ for the second line. \\

Combining Eqs.~\eqref{RB13} - \eqref{RB15}, one can see that the maximum amplitude of the AIC wavefunction near $\mathbf p_0$ scales as
\begin{equation}
    \begin{split}
        \max_\phi |\psi(r,\phi)| \sim \frac{|\lambda\psi_0|}{r}\frac{||\nabla H_0(\mathbf{p}_0)||}{J(\mathbf{p}_0)} \propto \frac{1}{r\,S(\mathbf{p}_0)}.
    \end{split}
\end{equation}

\section{Appendix V: Technical details of Eq.~(10),(11)in the maintext }
\subsection{Discussion on Lippmann-Schwinger formalism in non-Hermitian systems}

Lippmann-Schwinger formalism still holds in non-Hermitian systems, and thus Eq.~(10) is valid in non-Hermitian systems. 
To be precise, consider the eigenvalue equation 
\begin{equation}
    (E-\hat H_0-\hat V) \ket{\psi}=0
\end{equation}
This can be written as
\begin{equation}
    (E-\hat H_0)|\psi\rangle=\hat V|\psi\rangle
\end{equation}
The general solution is
a sum of a particular solution and a homogeneous solution:
\begin{equation}
|\psi\rangle = |\phi\rangle + (E-\hat H_0)^{-1}\hat V|\psi\rangle,
\label{eq:RC14}
\end{equation}
where $|\phi\rangle\in\ker(E-\hat H_0)$ is an eigenvector of
\(\hat H_0\) with eigenvalue $E$.
And this is exacly Eq.(10) in the maintext.

As a mathematical disclaimer,since $E$ is in the spectrum of $\hat{H}_0$, the operator $(E-\hat{H}_0)$ is not invertible (it has a kernel). Its position-space matrix elements, $ G(E,\mathbf{r},\mathbf{r}'):=\bra{\mathbf{r}}(E-\hat{H}_0)^{-1}\ket{\mathbf{r}'} $, is still perfectly well-defined. As we proved in the main text, this integral converges to a finite, algebraically decaying function,$|\mathbf{r}-\mathbf{r}'|^{-1}$ ,because the set of poles $S(E)$ (Eq.~(1) in the maintext) is discrete (measure zero) in 2D. In Hermitian case, one more step is needed, i.e., specify the boundary value by introducing $\pm i0^+$. But in this non-Hermitian case, we don't need this presciption since the matrix element is already finite and continuous with respect to $E$.   

Explicitly, in standard Hermitian scattering theory, the Green's operators $G^{\pm}(E)$ are defined as
\begin{equation}
    G^{\pm}(E) = \frac{1}{E - H_0 \pm i0^+},
\end{equation}
and the $\pm i0^+$ prescription is essential because the continuous spectrum lies on the real axis and produces a branch cut. The boundary values from above and below differ by a term proportional to $\delta(E-H_0)$; more precisely, the Sokhotski-Plemelj formula gives
\begin{equation}
    \frac{1}{E - h(\mathbf k) \pm i0^+}
 = \mathcal P\frac{1}{E-h(\mathbf k)} \mp i\pi \delta(E-h(\mathbf k)). \label{eq:Sokhotski}
\end{equation}
In our non-Hermitian setting the situation is fundamentally different.For energies $E$ inside the PBC continuum the matrix elements of the Green's operator,
\begin{equation}
     G_0(E;\mathbf r,\mathbf r')
 = \int_{\mathrm{BZ}}\frac{d^2\mathbf k}{(2\pi)^2}
 \frac{e^{i\mathbf k\cdot(\mathbf r-\mathbf r')}}{E-h(\mathbf k)}, 
\end{equation}
exist as convergent oscillatory integrals and are finite and continuous functions of $E$. The spectrum fills a two-dimensional region in the complex plane, so there is no 1D ``branch cut'' along which one has to take boundary values; consequently, the limits
\begin{equation}
 G_0^{+}(E) = \lim_{\eta\to0^+}G_0(E+i\eta),\qquad
 G_0^{-}(E) = \lim_{\eta\to0^+}G_0(E-i\eta)
 \end{equation}
coincide whenever these limits are defined. In other words, the retarded and advanced Green's operators agree. 

Physically, $\pm i0^+$ can be understood as the on-shell constraint during the scattering process, i.e., the $\delta$-term in Sokhotski-Plemelj formula(Eq.~\eqref{eq:Sokhotski}) enforces that only plane waves with energy $E$ contribute to the outgoing scattering wave, and this is what distinguishes ``in-going'' $(+)$ from ``out-going'' $(-)$ solutions. Thus, the scattering wave in Hermitian case is composed of plane waves with same energy $E$ as incoming wave and one typically interprets this result as that the incoming wave is scattered to other momentum with same energy~\cite{Economou2006}. 
However, in the non-Hermitian case, this on-shell constraint no longer holds. The scattering, as given by the Green's function, is an integral over the entire Brillouin zone. As we proved in our work, the far-field behavior of this wavefunction is dominated by the discrete singular $\mathbf{k}_0 \in S(E)$ (where $E=h(\mathbf{k}_0)$), and it is the local k-space structure around these $\mathbf{k}_0$ (as described in Eq.(4) in the maintext) that dictates the algebraic decay of the wavefunction. 

With this understanding, Eq.~\eqref{eq:RC14} still gives general solution of the Schrodinger equation by intepreting it as integral equation, 
\begin{equation}
    \psi(\mathbf{r}) = \phi(\mathbf{r}) + \int d\mathbf{r}'  G_0(E,\mathbf{r},\mathbf{r}') V(\mathbf{r}') \psi(\mathbf{r}'),
\end{equation}

\subsection{Subtley in degenerate case}
Though for each plane wave $\ket{\mathbf{k}}$, one can find corresponding AIC part, such AIC part is the same if two (or more) planes waves degenerate. To be more precise, for $E$ in the continuum spectrum, suppose that the original corresponding eigenspace of $\hat{H}_0$ is spanned by $\{\ket{\mathbf{k}_1}, ..., \ket{\mathbf{k}_s}\}$. Then the perturbed eigenspace of $\hat{H}$ is spanned by $\{\ket{\mathbf{k}_1}+\ket{AIC(E)}, ..., \ket{\mathbf{k}_s}+\ket{AIC(E)}\}$,i.e., same $\ket{AIC}=\hat{G}_0\ket{\mathbf{0}}$ part since the energy degenerates. Thus, one can also form eigenspace of $\hat{H}$ by substraction and getting $s-1$ linearly independent trivial states without AIC part and one states with AIC part,i.e., 
\begin{equation}
    \begin{split}
        &\text{span}\{\ket{\mathbf{k}_1}+\ket{AIC},\ket{\mathbf{k}_2}+\ket{AIC},...,\ket{\mathbf{k}_{s-1}}+\ket{AIC}, \ket{\mathbf{k}_s}+\ket{AIC}\} \\ 
        =&\text{span}\{\ket{\mathbf{k}_1}-\ket{\mathbf{k}_s},\ket{\mathbf{k}_2}-\ket{\mathbf{k}_s} ,...,\ket{\mathbf{k}_{s-1}}-\ket{\mathbf{k}_s}, \ket{\mathbf{k}_s}+\ket{AIC}\}
    \end{split}
\end{equation}

\section{Appendix VI: Verification on AIC as scattering wave in finite lattice}
To be specific, let us reserve the notation without superscript, $\hat{G}_0(E)$, for the Green's function in the thermodynamic limit. 
And we explicitly use $\hat{G}_0^L(E)$ for the one in an $L\times L$ finite lattice,

From discussions in the main text,  elements in the thermodynamic limit, $G_0(E,\mathbf{x})$, decays algebraically as a function of $\mathbf{x}$ . The element of $\hat{G}_0^L(E)$ can be expanded as
\begin{equation}
	\begin{split}
		G_0^L(E,\mathbf{x}) := \bra{\mathbf{x}}\hat{G}_0^L(E)\ket{\mathbf{0}} = \frac{1}{L^2}\sum_{\mathbf{k}\in BZ} \frac{e^{i \mathbf{k}\cdot\mathbf{x}}}{E-H_0(\mathbf{k})}. \label{eq:finite_green}
	\end{split}
\end{equation}

In finite lattice, all the $L^2$ perturbed eigenstates can be solved by the $L^2$-order algebraic equation, $f_\lambda(E)=0$, with $\hat{G}_0$ replaced by $\hat{G}_0^L$, since the spectrum is now discretized and isolated. 
We denote those eigenvalues as $\{E_i^\lambda\}_{i=1}^{L^2}$ , and at those points, the element of finite lattice Green's function, $G_0^L(E,\mathbf{0})$, differs from that of the thermodynamic limit, $G_0(E,\mathbf{0})$, by a finite amount, since the former is fixed to $\lambda^{-1}$ by the constraint and the latter varies continuously with respect to $E$. 

This deviation can be understood by observing that in Eq.~\eqref{eq:finite_green}, when $E_i^\lambda$ is near a singularity, $E_i^0 = H(\mathbf{k}_i)$,  the term $1/(L^2 (E_i^\lambda-E_i^0))$ can not be smoothed out and thus can not be approximated by the integration value at the thermodynamic limit. Separate such term and the remaining terms in Eq.~\eqref{eq:finite_green}, and denote $\Delta E_i = E_i^\lambda - E_i^0$ as the gap, 

\begin{equation}
	\begin{split}
		G_0^L(E,\mathbf{x}) &= L^{-2}\frac{e^{i \mathbf{k}_i\cdot \mathbf{x}}}{\Delta E_i} +L^{-2} \sum_{E_j\neq E_i}\frac{e^{i \mathbf{k}_j\cdot \mathbf{x}}}{E_i+ \Delta E_i - E_j} \\
		& \approx c_i e^{i \mathbf{k}_i\cdot \mathbf{x}} +G_0(E,\mathbf{x}), \label{eq:lambda_reformulated}
	\end{split}
\end{equation}
where $c_i = L^{-2}\Delta E_i^{-1}$. By taking $\mathbf{x}=0$, and noting that $G_0^L(E,\mathbf{0})= \lambda^{-1}$, we have $c_i = f_\lambda(E_i^\lambda)$ and thus Eq.~(11) in the main text is now proved. 

We remark that Eq.~\eqref{eq:lambda_reformulated} assumes no degeneracy for simplicity, since   finite-degeneracy complicates the situation only by introducing finitely more plane wave components with same coefficient, i.e., $c_i=c_j$ if $E_i^0=E_j^0$. However, above discussion fails for Hermitian case, since the degeneracy is infinite,implying the necessity of the discreteness of the singularity set $\{\mathbf{k}_0 \mid E_0 - h(\mathbf{k}_0) = 0\}$. 

\section{Appendix VII: Generalization to other potential profiles}

The algebraic localization in the maintext is in fact a universal feature of non-Hermitian systems, not an artifact of the delta potential. 

Physically, this generality stems from two core properties: (i) the continuum spectrum spans a finite area in the complex energy plane, and (ii) the disruption of translational symmetry induces localization in the bulk. Both of these properties hold for realistic potentials, provided their tails decay sufficiently fast (specifically, faster than $r^{-2}$, as we demonstrate below).
Rigorously, in the Green's function formalism, this algebraic localization is governed by the behavior of the unperturbed Green's function, $G_0(E,\mathbf{r}-\mathbf{r'})=\bra{\mathbf{r}}\hat{G}_0(E)\ket{\mathbf{r'}}$. We discuss this point in more detail below.

Let's denote the potential as $\hat{V}$ with profile in real space $\bra{\mathbf{r}}\hat{V}\ket{\mathbf{r}'} = \delta(\mathbf{r-r'}) V(\mathbf{r})$. The unperturbed Hamiltonian is $\hat{H}_0$ and the unperturbed Green's function is $\hat{G}_0(E)=(E-\hat{H}_0)^{-1}$. Recall the Schr\"odinger equation for a localized state in the continuum spectrum,
\begin{equation}
    (\hat{H}_0 + \hat{V})\ket{\psi} = E \ket{\psi}.
\end{equation}

Some algebraic manipulations lead to the Lippmann-Schwinger equation,
\begin{equation}
\psi(\mathbf{r}) = \int d\mathbf{r}'  \bra{\mathbf{r}}\hat{G}_0(E)\ket{\mathbf{r}'}  V(\mathbf{r}') \psi(\mathbf{r}').
\end{equation}
From the maintext, $G_0(E, \mathbf{r}-\mathbf{r}') := \bra{\mathbf{r}}\hat{G}_0(E)\ket{\mathbf{r}'}$ decays algebraically as $1/|\mathbf{r}-\mathbf{r}'|$ when $|\mathbf{r}-\mathbf{r}'| \to \infty$. 

In the following, we will prove that if the profile $V(\mathbf{r})$ decays sufficiently fast, the wavefunction $\psi(\mathbf{r})$ still decays algebraically as in the delta potential case.  The proof proceeds in two steps: first, we show this holds for $V(\mathbf{r})$ concentrated on a finite region, and second, we generalize the result to potentials with longer-range tails, such as Gaussian or screened Coulomb potentials.

\textbf{Step 1: potentials concentrated on a finite region}

Intuitively, for$|\mathbf{r}| \gg 1$, the integral is dominated by the region where $\mathbf{r}'$ is finite (i.e., the support of V). In this regime, $|\mathbf{r}-\mathbf{r}'| \approx |\mathbf{r}|$, which implies the algebraic decay. The other part of the integrand, $V(\mathbf{r}')\psi(\mathbf{r}')$, is assumed to be bounded within the finite support of the potential. This term can therefore be controlled by its supremum. 

To be concrete, let's denote $B(\mathbf{r},\xi)$ as a ball of radius $\xi$ centered at $\mathbf{r}$, i.e., 
$$
B(\mathbf{r},\xi):= \{\mathbf{r}': |\mathbf{r'- r}|<\xi\}.
$$
Suppose that $\hat{V}$ is supported in $B(\mathbf{0},\xi)$(i.e., $V(\mathbf{r}) = 0$ for $|\mathbf{r}|> \xi $). A crude bound yields

\begin{equation}
|\psi(\mathbf{r})| \le \sup_{|\mathbf{r}'|<\xi} |V(\mathbf{r}')\psi(\mathbf{r}')| \int_{|\mathbf{r}'|<\xi} d\mathbf{r}'  \frac{C}{|\mathbf{r}-\mathbf{r}'|}\le \sup_{|\mathbf{r}'|<\xi} |V(\mathbf{r}')\psi(\mathbf{r}')|\frac{C\pi \xi^2}{|\mathbf{r}|-\xi}\sim O(1/|\mathbf{r}|).  \label{eq:compact_estimation}
\end{equation}

\textbf{Step 2: potentials with extended but sufficiently fast decaying tails}

The method above can be generalized to potentials with longer but still sufficiently fast decaying tails, such as Gaussian potentials or screened Coulomb potentials.  The idea used here is to divide the integration region into a near region and tail region. The near region can be estimated as step 1. The tail region can be controlled both by the decay of $V(\mathbf{r})$ (for contribution near the observing point $\mathbf{r}$) and the algebraic decay of $G_0(E,\mathbf{r}-\mathbf{r}')$. 

Let $\xi >0$ be the corresponding characteristic length scale. Then Eq.~\eqref{eq:compact_estimation} estimates the contribution from the region $B(\mathbf{0},\xi)$. 

For the tail region ,we divide it into two parts: $B(\mathbf{0},\xi)^c \cap B(\mathbf{r},\frac{|\mathbf{r}|}{2})$ and $B(\mathbf{0},\xi)^c \cap B(\mathbf{r},\frac{|\mathbf{r}|}{2})^c$, where $A^c$ is the complement of set $A$.
 
The first part is the contribution near the observation point $\mathbf{r}$.We assume $V(\mathbf{r})$ is spherically symmetric ($V(\mathbf{r})=V(|\mathbf{r}|)$) and monotonically decreasing. In this region, $|\mathbf{r}'| \ge |\mathbf{r}|/2$, so $V(|\mathbf{r}'|) \le V(|\mathbf{r}|/2)$. Bounding the integral yields

\begin{equation}
    \int_{B(\mathbf{r},\frac{|\mathbf{r}|}{2})} d\mathbf{r}' |G_0(E,\mathbf{r}-\mathbf{r}')|  |V(\mathbf{r}') \psi(\mathbf{r}')|
    \le (\sup |\psi|) V(\frac{|\mathbf{r}|}{2}) \int_{B(\mathbf{r},\frac{|\mathbf{r}|}{2})} d\mathbf{r}' \frac{C}{|\mathbf{r}-\mathbf{r}'|} .\label{eq:tail1}
\end{equation}

By defining $\mathbf{u}= \mathbf{r}-\mathbf{r}'$ and changing to polar coordinates, the remaining 2D integral in Eq.~\eqref{eq:tail1} is 
\begin{equation}
    \int_{B(\mathbf{0},\frac{|\mathbf{r}|}{2})} d^2\mathbf{u} \frac{C}{|\mathbf{u}|} = C \int_0^{2\pi}d\theta \int_0^{\frac{|\mathbf{r}|}{2}} \frac{1}{u} udu = \pi C |\mathbf{r}|.
\end{equation}
Thus, the bound for this part is $O(V(\frac{|\mathbf{r}|}{2}) |\mathbf{r}|)$. 

For the remaining part, one has
\begin{equation}
    \begin{split}
    \int_{B(\mathbf{0},\xi)^c \cap B(\mathbf{r},\frac{|\mathbf{r}|}{2})^c} d\mathbf{r}' |G_0(E,\mathbf{r}-\mathbf{r}')|  |V(\mathbf{r}') \psi(\mathbf{r}')|
    &\le (\sup \psi(\mathbf{r}')|) \int_{|\mathbf{r}'|\ge \xi, |\mathbf{r}'-\mathbf{r}|>\frac{|\mathbf{r}|}{2}|} d\mathbf{r}'  \frac{C}{|\mathbf{r}'-\mathbf{r}|}|V(\mathbf{r}')|.\\
    & \le \frac{2C(\sup \psi(\mathbf{r}')|)}{|\mathbf{r}|}  \int_{|\mathbf{r}'|\ge \xi} d\mathbf{r}' |V(\mathbf{r}')| \\
    & \sim O(1/|\mathbf{r}|). \label{eq:tail2}
    \end{split}
\end{equation}
Combining Eqs.~\eqref{eq:compact_estimation}, \eqref{eq:tail1}, and \eqref{eq:tail2}, we see that all contributions are  $O(|\mathbf{r}|^{-1})$ or smaller provided $V$ decays faster than $r^{-2}$. 

Concretely, bound of Eq.~\eqref{eq:tail1} requires $O(V(\frac{|\mathbf{r}|}{2}) |\mathbf{r}|)\le O(|\mathbf{r}|^{-1})$ , which requires $V(\mathbf{r})$ to decay faster than $r^{-2}$. The bound of Eq.~\eqref{eq:tail2} holds only if $\int_{|\mathbf{r}'|\ge \xi} d\mathbf{r}' |V(\mathbf{r}')|$ converges, which also require $V(\mathbf{r})$ to decay faster than $r^{-2}$. 

Therefore, the algebraic decay $O(|\mathbf{r}|^{-1})$ holds as long as $V(\mathbf{r})$ decays faster than $r^{-2}$, which is true for Gaussian and screened Coulomb potentials. 

\textbf{Energy of AICs and Numerical simulations}
\begin{figure*}[t]
	\begin{centering}
		\includegraphics[width=1\linewidth]{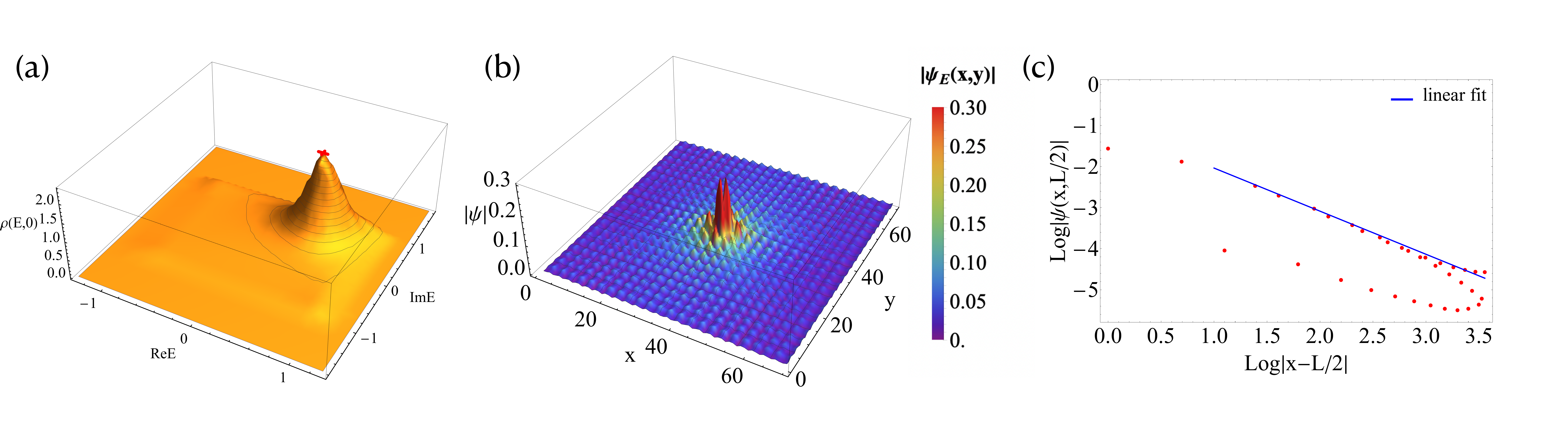}
		\par\end{centering}
		\protect\caption{
		(a) LDOS at the impurity site for Gaussian potential with $V(\mathbf{r})=\lambda  e^{-|\mathbf{r}|^2/2\sigma^2}$, where $\sigma=0.3$ and $\lambda\approx 3.0882 (1+ i)$. The Hamiltonian is chosen to be $h(k_x,k_y)= \cos k_x + i \cos k_y$ with lattice size $70 \times 70$. The red cross($E\approx 0.48 + 0.52i$) marks the peak of LDOS inside the bulk continuum spectrum, which corresponds to an AIC. 
        (b) The corresponding wavefunction profile for the peak in (a).
        (c) The log-log plot of wavefunction amplitude along $x$-axis, where $x$ ranges from $L/2$ to $L$,i.e.,right half of (b), which shows clear algebraic decay. The fitted slope is $k\approx -1.05$.
		}\label{fig:smf2}
\end{figure*}

The energy of AICs, which is harder to determine analytically for longer-range potentials, can be found numerically or experimentally by identifying peaks in the LDOS at the impurity site, as introduced in the main text. Concretely, similar to the delta potential case in the main text, the general states in the continuum can be expressed via the Lippmann-Schwinger equation:
\begin{equation}
	\begin{split}
		\ket{\psi} = \ket{\phi}+  (E-\hat{H}_0)^{-1} \hat{V}\ket{\psi} ,\label{eq:LS_eq}
	\end{split}
\end{equation}
where $\ket{\phi} \in \ker\{E-\hat{H}_0\}$ is a general eigenstate of $H_0$ with energy $E$. At the specific AIC energy, the homogeneous part $\ket{\phi}$ vanishes(i.e.,$\ket{\phi}=0$), and the total eigenstate $\ket{\psi}$ is purely the localized state. Thus, near this energy, the non-zero $\ket{\phi}$component makes the state less localized at the impurity, causing the peak in the LDOS which serves as a clear signature. \\

In Fig.~\ref{fig:smf2}(a), we provide the numerical result for the LDOS at the impurity site for the Gaussian potential. One can see a clear peak (red cross) inside the bulk continuum spectrum, which corresponds to the AIC. The corresponding wavefunction profile at the peak energy is shown in Fig.~\ref{fig:smf2}(b), which shows clear $O(r^{-1})$ algebraic decay(Fig.~\ref{fig:smf2}(c)).

\section{Appendix VIII: Discussion on open boundary case}
\subsection{Corner skin case}
To be specific, let's consider the model 
\begin{equation}
    h(\mathbf{k}) = t_r e^{ik_x}+ t_l e^{-ik_x} + t_d e^{ik_y} + t_u e^{-ik_y}. \label{eq:B_skin_model}
\end{equation}
Without loss of generality, let's choose $|t_r| > |t_l|$ and $|t_d| > |t_u|$, so that skin modes accumulate at the bottom-left corner under OBC.\\

The GBZ of Eq.~\eqref{eq:B_skin_model} can be decomposed by $\text{GBZ}=\text{GBZ}_x \otimes \text{GBZ}_y$, where $\text{GBZ}_{x(y)}= \{e^{\mu_{x(y)} + i k}| k\in (0,2\pi]\}$ for $\mu_x =\frac{1}{2} \ln|\frac{t_l}{t_r}| $ and $\mu_y =\frac{1}{2} \ln|\frac{t_u}{t_d}| $.\\

Fix a reference energy $E$ inside OBC continuum spectrum and let's consider the spatial profile of the corresponding exited state, which is propotional to $G_0(E,\mathbf{r},\mathbf{r_0})=\bra{\mathbf{r}}\hat{G}_0(E)\ket{\mathbf{r_0}}$, where $\mathbf{r_0}$ stands for the position of impurity, and $\hat{G}_0(E)= (E-\hat{H}_0)^{-1}$ for unperturbed Green's function under corresponding boundary condition.

\begin{itemize}
    \item full OBC 
\end{itemize}
Decompose the Green's function with the GBZ basis, one has
\begin{equation}
    \begin{split}
        G_0(E,\mathbf{r},\mathbf{r_0}) =\frac{1}{(2\pi)^2} \int_{0}^{2\pi}\int_{0}^{2\pi}d^2 \mathbf{k}  \frac{e^{(\boldsymbol{\mu}+i\mathbf{k})\cdot (\mathbf{r}-\mathbf{r_0})}}{E-h(-i\boldsymbol{\mu}+\mathbf{k})} \;\; , \boldsymbol{\mu}= (\mu_x,\mu_y). 
    \end{split}
\end{equation}
Separating overal exponential $e^{\boldsymbol{\mu}\cdot (\mathbf{r}-\mathbf{r_0})}$ factor, the remaining part of the integral is the same as AICs for model
\begin{equation}
    \tilde{h} (\mathbf{k}) = (t_r e^{\mu_x}) e^{ik_x}+ (t_l e^{-\mu_x}) e^{-ik_x} + (t_d e^{\mu_y}) e^{ik_y} + (t_u e^{-\mu_y}) e^{-ik_y},
\end{equation}
under PBC condition, which behaves like $O(|\mathbf{r}-\mathbf{r_0}|^{-1})$. Thus, the wavefunction behaves like
\begin{equation}
    \begin{split}
        \psi(\mathbf{r}) \sim O(|\mathbf{r}-\mathbf{r_0}|^{-1} e^{-\boldsymbol{\mu}\cdot \mathbf{r}}),
    \end{split}
\end{equation}
where the overal constant $e^{-\boldsymbol{\mu}\cdot \mathbf{r_0}}$ can be removed by normalization. Thus, the algebraic localization is greatly suppressed by the exponential decay from skin effect, and this suppression is not influenced by impurity position $\mathbf{r}_0$.

\begin{itemize}
    \item hybrid: OBC at $x$-direction + PBC at $y$-direction
\end{itemize}
Similarly with full OBC case, one has overall exponential decay $e^{-\mu_x (x-x_0)}$ from skin effect along $x$-direction, and the remaining integral is the same as AICs, which gives the algebraic decay. Thus, the wavefunction behaves like
\begin{equation}
    \begin{split}
        \psi(\mathbf{r}) \sim O(|\mathbf{r}-\mathbf{r_0}|^{-1} e^{-\mu_x x}).
    \end{split}
\end{equation}
Thus, a impurity at the boundary $r_{0,x}=0$ makes the algebraic localization along $y$ direction more significant and easier to observe, while an impurity deep in the bulk/other side of the lattice makes the algebraic localization along $y$ direction harder to observe because of the suppression of skin effect along $x$-axis.

\subsection{Algebraic skin case}
In the presence of an algebraically localized skin effect, the analysis of AICs becomes more subtle. In contrast to the corner-skin case, the Generalized Brillouin Zone (GBZ) is not uniquely defined and the biorthogonal mode expansion is not available in a simple closed form~\cite{Zhang2025}. For this reason we cannot yet give a fully rigorous statement. Nevertheless, our physical considerations and numerics both suggest that AICs created by a bulk impurity are strongly suppressed and practically unobservable in this regime.

Intuitively, an AIC is a collective state built from unperturbed bulk eigenstates in a narrow neighborhood of a Bloch momentum $\mathbf p_0$ satisfying $E = h(\mathbf p_0)$. In the PBC continuum analysis of the main text, this shows up as the fact that the far-field profile of the AIC is controlled by the linearized dispersion $h(\mathbf p_0 + \delta\mathbf p)\simeq E + \mathbf v\cdot\delta\mathbf p$, with group velocity $\mathbf v = \nabla_{\mathbf p} h(\mathbf p_0)$. In an algebraic-skin system under OBC, however, the relevant OBC eigenstates are already strongly biased toward the boundary: their weight at a site deep in the bulk (such as the impurity position $\mathbf r_0$) scales down algebraically with the distance to the boundary. Constructing an impurity-centered localized profile from such boundary-localized modes therefore requires a finely balanced superposition of many states whose individual amplitudes at $\mathbf r_0$ are parametrically small, so the resulting impurity effect is expected to be very weak. 

Moreover, the algebraic skin modes themselves are a robust consequence of the OBC geometry~\cite{Zhang2025}: they are fixed by the global boundary conditions and the generalized Fermi-surface structure of the bulk band, and are not easily deformed by a single local impurity in the interior. This further supports the expectation that a bulk impurity will only weakly perturb the existing algebraic skin profile rather than generate a pronounced new impurity-localized state.

\begin{figure*}[t]
	\begin{centering}
		\includegraphics[width=1\linewidth]{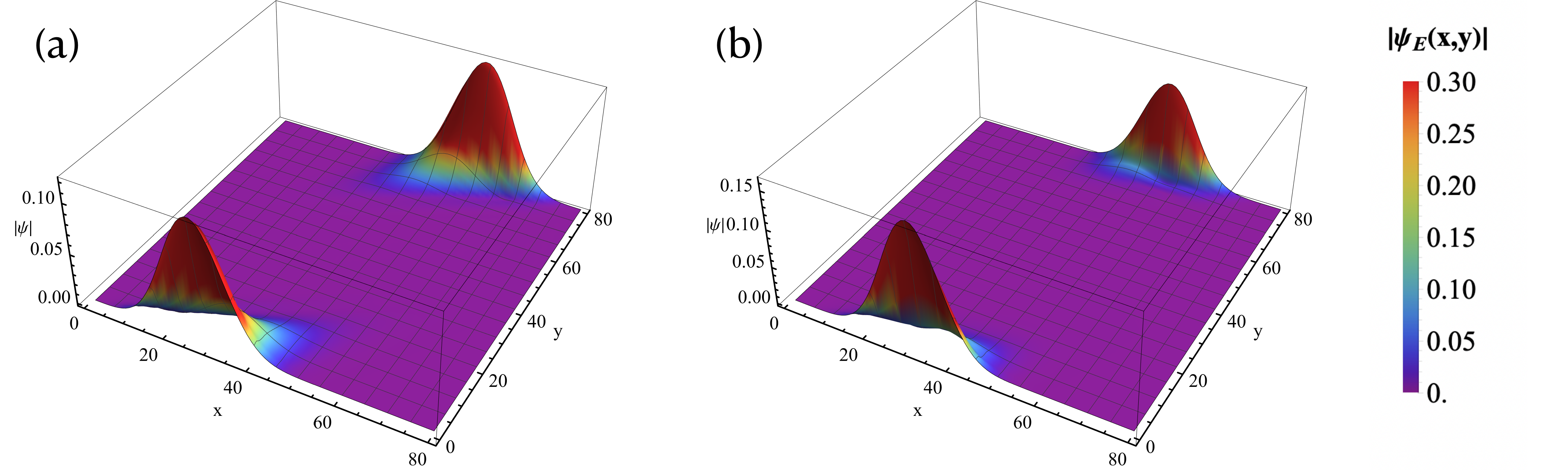}
		\par\end{centering}
		\protect\caption{
Spatial profile of algebraic skin modes in the presence of a bulk impurity.
  We use the reciprocal model $h(\beta_x,\beta_y)
  = i(\beta_y + \beta_y^{-1}) + \beta_x\beta_y + (\beta_x\beta_y)^{-1}$ 
  on an $80\times 80$ lattice, which exhibits algebraic skin modes localized near the $x$ boundaries~\cite{Zhang2025}. 
  A single on-site impurity is placed at the center of the lattice.
  Representative eigenstates for (a) $E = 1 + 1.5i$, $\lambda = 3 + 2i$ and 
  (b) $E = 1 + 1.2i$, $\lambda = 0.3 + 0.2i$. 
  In both cases, the overall algebraic skin profile remains essentially unchanged and no additional impurity-centered peak is visible.
}\label{fig:smf3}
\end{figure*}

The numerical observations (Fig~\ref{fig:smf3}) are consistent with the above picture: for typical parameters in the algebraic-skin regime, the impurity introduces only tiny corrections to the algebraic skin modes, and we do not observe a distinct AIC-type peak localized at the impurity.

\section{Appendix IX: Generalization to arbitrary $d>1$ dimensions}
Intuitively, AICs can be understood as a consequence of area-typed spectrum and disruption of translational invariance in the bulk, both of which still hold in higher dimensions. Thus, it can be expected that AIC still exists in higher dimensions, which is indeed the case. In the following, we give a detailed generalization in higher dimensions. Actually, the dimension only changes the order of the algebraic decay: similar to $r^{-1}$ decay in two dimensions, the wave function decays as $r^{-\frac{d}{2}}$ generally with the exception of $r^{-1}$ for specific directions in $d>1$ dimensions. Below we give a detailed arbitrary $d>1$ generalization beginning with $d=3$ case. 

\subsection{Solvable toy continuum model}  

Let's begin with a toy continuum model in $3d$ case ($c>0$)
\begin{equation}
    (p_x^2 + i(p_y^2+p_z^2 - c^2) -  \delta(x,y,z))\psi(x,y,z) = 0 \label{eq:3d_continuum_schordinger}
\end{equation}
and with Fourier transformation, one has(for compactness of notation, we've rescaled $\psi$ to $\frac{\psi}{\psi(\mathbf{r}=\mathbf{0})}$).
\begin{equation}
    \psi(\mathbf{r} )= \int d^3 \mathbf{k}  \frac{e^{i \mathbf{k\cdot r}}}{k_x + i (k_y^2 + k_z^2 -c^2)} \label{eq:3d_continuum}
\end{equation}

and we move to cylindrical coordinate : $\mathbf{k}= (k_x, k ,\theta), \mathbf{r}=(x, r,\alpha)$ . Without loss of generality, let's assume $x>0$ so that we close the $k_x$ integration with upper half loop to use residue theorem
\begin{equation}
    \psi(\mathbf{r}) = \int_0^c kdk \int_0^{2\pi}d\theta  e^{(k^2-c^2)x} e^{ikr\cos\theta}.
\end{equation}
Thus , along $x$-axis
\begin{equation}
    \begin{split}
        \psi(x,0,0) &= 2\pi\int_0^c k e^{(k^2 -c^2)x}dk  \\
        &= 2\pi e^{-c^2x}\frac{e^{c^2x} -1}{2x}\\
        &\sim O(x^{-1}).
    \end{split}
\end{equation}

In the $(y,z)$ plane,
\begin{equation}
    \begin{split}
        \psi(x=0,r) &= \int_0^c kdk \int_0^{2\pi}d\theta   e^{ikr\cos\theta} \\
&= \int_0^c k J_0(kr)dk \\
&= \frac{1}{r^2} \int_0^{cr}xJ_0(x)dx\\
&= r^{-2} crJ_1(cr)
    \end{split}
\end{equation}

Since $J_1(r)\sim r^{-\frac{1}{2}},r\to \infty$, one has
\begin{equation}
    \psi(x=0,r)\sim O(r^{-\frac{3}{2}}).
\end{equation}

Below, we shall show $O(r^{-\frac{3}{2}})$ is general.

\subsection{General case}

Suppose that the one-dimensional zero-locus is 
\begin{equation}
    \mathcal{C}_E := \{\mathbf{k} | E-h(\mathbf{k})=0\}
\end{equation}
and we use $s$ to parametrize it (say, arc-length). At $\mathbf{k}_0(s)\in \mathcal{C}_E$, we build the local frame,
\begin{equation}
    \{\mathbf{t,n_1,n_2}\}(s),
\end{equation}
where $\mathbf{t}(s)$ is the tangent vector, and $\mathbf{n}_{1,2}$ span the normal plane. 

Generally for $\mathbf{k}$ near $\mathcal{C}_E$, one has
\begin{equation}
    \begin{split}
       \mathbf{k} &= \mathbf{k}_0(s) + u_1 \mathbf{n}_1 + u_2 \mathbf{n}_2 \\
h(\mathbf{k} ) &\approx E+ \alpha_1 u_1 + \alpha_2 u_2 , \alpha_{1,2}\in \mathbb{C}. \label{eq:sub48}
    \end{split}
\end{equation}

And thus, 
\begin{equation}
    \begin{split}
        \psi(\mathbf{r}) = \int ds  e^{i \mathbf{k}_0(s)\cdot \mathbf{r}}  
\int du_1 du_2 \frac{e^{i(\eta_1 u_1 + \eta_2 u_2)r}}{-( \alpha_1 u_1 + \alpha_2 u_2)} \label{eq:sub49}
    \end{split}
\end{equation}

The ($u_1,u_2$) integration give $r^{-1}$ factor as in two dimension case. And from stationary-phase approximation,  $s$ integration gives $r^{-\frac{1}{2}}$ factor.

Thus, generally, one has
\begin{equation}
    \psi(\mathbf{r}) \sim O(r^{-\frac{3}{2}})
\end{equation}

\subsection{Revisite of toy model}

Now, let's explicitly construct above parametrization with the continuum model Eq.~(\ref{eq:3d_continuum}).

The zero set is the circle
\begin{equation}
    \mathcal{C}_E = \{\mathbf k_0(s)=(0,c\sin s,c\cos s),s\in[0,2\pi)\},
\end{equation}

Use the moving orthonormal frame at $\mathbf k_0(s)$:
\begin{equation}
    \mathbf t(s)=(0,-\sin s,\cos s),\qquad
\mathbf n_1=(1,0,0),\qquad
\mathbf n_2(s)=(0,\cos s,\sin s).
\end{equation}

Thus, for a general $\mathbf{k}$ near $\mathcal{C}_E$, the decompostion becomes
\begin{equation}
    \begin{split}
        s(\mathbf{k})&=\arctan(\frac{k_y}{k_z}), \\
u_1(\mathbf{k})&=  (\mathbf{k}-\mathbf{k}_0(s))\cdot \mathbf{n}_1 = k_x, \\
u_2(\mathbf{k})&=  (\mathbf{k}-\mathbf{k}_0(s))\cdot \mathbf{n}_2 = \sqrt{k_y^2+ k_z^2}-c
    \end{split}
\end{equation}

Finally, the denominator and exponent in the numerator becomes
\begin{equation}
    \begin{split}
        k_x^2 + i(k_y^2+ k_z^2 -c^2) &\approx u_1 + 2 ic u_2,  \\
\mathbf{k}\cdot \mathbf{r}&=\mathbf{k}_0(s)\cdot \mathbf{r} + r(\eta_1 u_1 + \eta_2 u_2),
    \end{split}
\end{equation}

where $\mathbf{r}=r(r_x,r_y,r_z)$, and $\eta_1 = r_x, \eta_2 = r_y \cos s+ r_z \sin s$.  

Thus
\begin{equation}
    \psi(\mathbf{r}) \approx \int ds e^{i\mathbf{k}_0(s)\cdot \mathbf{r}}\int du_1 du_2 \frac{e^{ir(\eta_1 u_1 + \eta_2 u_2)}}{u_1 + 2 ic u_2}.
\end{equation}

And one can conclude that generally $\psi(\mathbf{r})\sim O(r^{-\frac{3}{2}})$ except along x-axis, where $\mathbf{k}_0(s)\cdot \mathbf{r}\equiv 0$ and stationary phase approximation no longer holds. Geometrically, this direction is special because it's perpendicular to the plane where the zero set lies. 

\subsection{ General higher dimensions} 

Based on above discussion for $d=3$ case, one can show that for general $d>1$, AICs decays algebraically as $r^{-\frac{d}{2}}$ with $r^{-1}$ as exception for specific directions.

Generaly, for $d-$dimensional systems, the zero set, $\mathcal{C}_E$, is a $(d-2)-$dimensional curve. Denote $\mathbf{s}=(s_1,s_2,..,s_{(d-2)})$ for the local coordinates of $\mathcal{C}_E$ and $(u_1,u_2)$ for the remaining dimensions, i.e., coordinates for the neighborhood of $\mathcal{C}_E$ with $\{\mathbf{t_1,...,t_{(d-2)},n_1,n_2}\}(\mathbf{s})$ as corresponding unit vector. Then Eq.~\ref{eq:sub48} still holds with modification of replacing scalar $s$ by $(d-2)-$dimensional coordinates $\mathbf{s}$. Thus,  Eq.~\ref{eq:sub49} can be generalized to  
\begin{equation}
    \begin{split}
         \psi(\mathbf{r}) = \int ds_1 ...ds_{(d-2)}  e^{i \mathbf{k}_0(\mathbf{s})\cdot \mathbf{r}}  
\int du_1 du_2 \frac{e^{i(\eta_1 u_1 + \eta_2 u_2)r}}{-( \alpha_1 u_1 + \alpha_2 u_2)} . 
    \end{split}
\end{equation}

In above integration, $(u_1,u_2)$ integral part gives $r^{-1}$ decay as in $d=2,3$ case. And from stationary phase approximation, the asymptotics of $\mathbf{s}$ integral part is $r^{-\frac{d-2}{2}}$. Combining those two parts together, $\psi(\mathbf{r})\sim r^{-\frac{d}{2}}, r\to \infty$. The exception comes from the direction $\mathbf{r}$ where $\mathbf{k}_0(\mathbf{s})\cdot \mathbf{r}\equiv0$, and $\psi(\mathbf{r})$ decay as $r^{-1}$ for such direction.

\end{widetext}
\end{document}